\documentclass[aip,jmp,reprint,onecolumn,preprintnumbers]{revtex4-1}

\usepackage{amsmath}
\usepackage{amssymb}
\usepackage{amsfonts}
\usepackage{graphicx}
\usepackage{slashed}
\usepackage{calrsfs}
\usepackage{tikz}
\usetikzlibrary{arrows,decorations.pathmorphing,backgrounds,positioning,fit,matrix}

\usepackage[utf8]{inputenc}

\draft 

\begin{document}

\preprint{IPPP/16/94}

\title{Landau-Khalatnikov-Fradkin transformation\\ for the fermion propagator in QED in arbitrary dimensions} 



\author{Shaoyang Jia }
\email{sjia@email.wm.edu}
\affiliation{Physics Department, College of William \& Mary, Williamsburg, VA 23187, USA}
\author{ M.R. Pennington }
\email{michaelp@jlab.org}
\affiliation{Physics Department, College of William \& Mary, Williamsburg, VA 23187, USA}
\affiliation{Theory Center, Thomas Jefferson National Accelerator Facility, Newport News, VA 23606, USA}

\date{\today}

\begin{abstract}
We explore the dependence of 
fermion propagators on the covariant gauge fixing parameter in quantum electrodynamics (QED) with the number of spacetime dimensions kept explicit. Gauge covariance is controlled by the the Landau-Khalatnikov-Fradkin transformation (LKFT). Utilizing its group nature, the LKFT for a fermion propagator in Minkowski space is solved exactly. The special scenario of 3D is used to test claims made for general cases. When renormalized correctly, a simplification of the LKFT in 4D has been achieved with the help of fractional calculus. 
\end{abstract}


\maketitle 

	\section{Introduction}
	The covariant gauge fixing procedure of quantum electrodynamics (QED) introduces a parameter $\xi$ into its Lagrangian, as a result of which Green's functions have explicit dependence on this gauge fixing parameter \cite{peskin1995introduction}. How one specific Green's function changes from one gauge to another is given by the Landau-Khalatnikov-Fradkin transformation (LKFT) \cite{Landau:1955zz,Fradkin:1955jr,Zumino:1959wt}. For brane worlds this has been recently derived in Ahmad \textit{et al.}\cite{Ahmad:2016dsb}. The generalization to $\mathrm{SU(N)}$ gauge theories is given by Aslam \textit{et al.}\cite{Aslam:2015nia}. The inclusion of LKFT into truncation schemes for the Schwinger-Dyson equations (SDEs) of QED ensures solutions under such schemes are gauge covariant. While the SDE for the fermion propagator involves the fermion-photon three-point function, the LKFT only involves the fermion propagator itself, making it possible for exact solutions to be obtained without introducing an ansatz for the vertex. 
	The purpose of this article is to deduce exact solutions of LKFT, laying the foundation for further work on their consistency with SDEs.
	
	LKFT was originally formulated for Green's functions in coordinate space. In contrast, it is in momentum space that the fermion propagator has been solved extensively from its Schwinger-Dyson equation (SDE) \cite{Kizilersu:2014ela}. As a result, checking gauge covariance of these solutions can be achieved only by lengthy procedures of Fourier transforms \cite{Bashir:2002sp}. Meanwhile, the rich structure of a momentum space fermion propagator is more transparent in Minkowski space, especially in the timelike region. It is therefore highly desirable if the LKFT can be solved directly for the momentum space propagator in the timelike region. With reasonable assumptions on their analytic structures, propagator functions in Minkowski space can be elegantly described using spectral representations \cite{peskin1995introduction}. As we will see, this also allows easy application of perturbative techniques, including Feynman parameterization \cite{peskin1995introduction} and dimensional regularization \cite{'tHooft:1972fi}, into nonperturbative calculations.
	
	The spectral representation for a scalar propagator in momentum space \cite{peskin1995introduction} is given by ${D(p^2)=\int_{m^2}^{+\infty}ds~\rho(s)/(p^2-s+i\varepsilon)}$, provided that the function $D(p^2) \to 0$ when $p^2 \to \infty$ in all directions in the complex plane, otherwise subtractions may be required as discussed later. For a free particle, the spectral function is simply $\rho(s)=\delta(s-m^2)$. When interactions are present, real particles can be produced by quantum loop corrections, adding  a $\theta$-function term to the spectral function at each multiparticle threshold. The spectral function is defined in a broader sense than regular functions. If there is not a component in the propagator function $D(p^2)$ more singular when $p^2 \to m^2$ than the free-particle propagator, a $\delta$-function and a $\theta$-function are all the terms needed in $\rho(s)$. However, when $D(p^2)$ is allowed to be more singular, there will be more  terms in the distribution $\rho(s)$ that correspond to derivatives of a $\delta$-function.

	
	QED in 4D being renormalizable, its divergences  are best captured as long known by dimensional regularization \cite{'tHooft:1972fi} as this preserves gauge symmetry and translational invariance. Here we solve the LKFT for the gauge covariant behavior of fermion propagator using a  spectral representation. Consequently continuing in the number of spacetime dimensions provides a convenient way  to regularize behaviors more singular than free-particle propagators  at the real particle production thresholds. Moreover keeping the number of spacetime dimensions explicit also allows simultaneous calculation of results in 3D and 4D, both of which are of current interest. As we will see the LKFT for fermion propagator in 3D is simpler than that in 4D. Explicit solutions to LKFT in 3D will be used  to illustrate properties of the LKFT made in the general case. What is more the dependence of the solutions on $\epsilon=2-d/2$ provides insights into how gauge covariance of QED in different dimensions are connected explicitly.
	
	We take the the Bjorken-Drell metric such that $p^2>0$ corresponds to the timelike momentum while $p^2<0$ for spacelike. The anti-commutation relations for Dirac gamma matrices are $\{\gamma^\mu,\gamma^\nu\}=2g^{\mu\nu}$. To avoid confusion, the variation of Greek letter epsilon, written as $\varepsilon$ is used for the Feynman prescription, while $\epsilon$ is related to the number of spacetime dimensions by $d=4-2\epsilon$.
	
	This article is organized as follows. In Section \ref{ss:LKFT_group}, the LKFT for the momentum space fermion propagator is obtained. After a review of the spectral representation for propagators, we establish three isomorphic representations of the LKFT as a continuous group parametrized by the covariant gauge parameter $\xi$. In Section \ref{ss:LKFT_spectral_rep}, the LKFT for fermion propagator spectral functions has been solved exactly with the number of spacetime dimensions kept explicit. In Section \ref{ss:LKFT_fermion_3D}, group properties of the LKFT are tested in 3D. When renormalized correctly, the LKFT for fermion propagator spectral functions in 4D is presented in Section \ref{ss:LKFT_fermion_4D}. We give our final remarks in Section \ref{ss:conclusions}.
	\section{LKFT as group transforms\label{ss:LKFT_group}}
	\subsection{LKFT in differential form}
	To derive the LKFT for the QED fermion propagator in the momentum space, first consider that under two gauge fixing conditions the coordinate space photon propagator changes from $D_{\mu\nu}(z)$ to $D_{\mu\nu}'(z)$;
	\begin{equation}
	D_{\mu\nu}'(z)=D_{\mu\nu}(z)+\partial_\mu\partial_\nu \delta M(z).
	\end{equation}
	According to Zumino \cite{Zumino:1959wt}, coordinate space fermion propagators evaluated with the corresponding gauge fixing conditions are related by
	\begin{equation}
	S_F'(x-y)=\exp\big\{ie^2\left[\delta M(x-y)-\delta M(0)\right]\big\}S_F(x-y).\label{eq:LKFT_SF}
	\end{equation}
	Specifically for our interest, starting from the Landau gauge to any other covariant gauge, function $\delta M(z)$ becomes
	\begin{equation}
	\delta M(z)=\xi M(z)=-\xi \int d\underline{l}\;\frac{e^{-il\cdot z}}{l^4+i\epsilon},\label{eq:def_M_covariant}
	\end{equation}
	where $d\underline{l}$ denotes the $d$-dimensional momentum measure $d\underline{l}\equiv d^d l/(2\pi)^d$.
	Substituting \eqref{eq:def_M_covariant} into \eqref{eq:LKFT_SF} produces the LKFT for the covariant gauge fermion propagator in coordinate space. In principle, taking the Fourier transform of \eqref{eq:LKFT_SF} gives the LKFT for fermion propagators in momentum space. In practice this is difficult to accomplish because of the exponential factor in \eqref{eq:LKFT_SF} defined by $M(z)$ in \eqref{eq:def_M_covariant} remaining illusive. However, it has been shown that if the fermion propagator takes its free-particle form in the Landau gauge, the Fourier transform of \eqref{eq:LKFT_SF} can be calculated~\cite{Bashir:2002sp}.
	
	While we are interested in the scenario where the fermion propagator in the Landau gauge is more than the free-particle propagator, to circumvent the difficulty of performing Fourier transforms of implicit functions, consider taking a first order derivative with respect to $\xi$ of \eqref{eq:LKFT_SF}. Noting that $S_F(x-y)$ on the right of \eqref{eq:LKFT_SF} is in a specific gauge, we have the result,
	\begin{equation}
	\dfrac{\partial}{\partial\xi}S_F'(x-y)=ie^2[M(x-y)-M(0)]\,S_F'(x-y).\label{eq:Dxi_SF_cord}
	\end{equation}
	Notice that the exponential factor has been absorbed into the Landau gauge propagator using  \eqref{eq:LKFT_SF}, giving rise to the $S'_F(x-y)$ factor on the right-hand side of  \eqref{eq:Dxi_SF_cord}. Now that all propagators in  \eqref{eq:Dxi_SF_cord} are primed, the prime notation can be dropped. We use $S_F(p;\xi)$ to denote the propagator in momentum space in any covariant gauge. Since there are no implicit functions left, taking the Fourier transform of  \eqref{eq:Dxi_SF_cord} gives
	\begin{equation}
	\dfrac{\partial}{\partial\xi}S_F(p;\xi)=ie^2\int d\underline{l}\;\dfrac{1}{l^4+i\epsilon}\;[S_F(p;\xi)-S_F(p-l;\xi)].\label{eq:Dxi_SF}
	\end{equation}
	Equation~\eqref{eq:Dxi_SF} is the LKFT for the momentum space fermion propagator, in differential form. Unlike  \eqref{eq:LKFT_SF}, differentiating means there is no explicit dependence on the initial condition. 
	
	However, when the propagator goes to a constant while $p^2\rightarrow \infty$, the following rewriting might be required
	\begin{equation}
	S_F(p;\xi)=R_F(\xi)+\tilde{S}_F(p;\xi).\label{eq:S_j_subtraction}
	\end{equation}
	Since the Fourier transform of a constant is $\delta$-function,  \eqref{eq:S_j_subtraction} indicates, in the coordinate space,
	\begin{equation}
	S_F(x-y;\xi)=R_F(\xi)\delta(x-y)+\tilde{S}_F(x-y;\xi).
	\end{equation}
	Substituting it into  \eqref{eq:LKFT_SF} gives $R_F(\xi)=R_F(0)$ and
	\begin{equation}
	\tilde{S}_F(x-y;\xi)=\exp\big\{ie^2[\delta M(x-y)-\delta M(0)] \big\}\tilde{S}_F(x-y;0).
	\end{equation}
	Therefore the LKFT for the subtracted propagator is identical to  \eqref{eq:LKFT_SF}.
		
	As with any first order differential equations, solving for the fermion propagator from  \eqref{eq:Dxi_SF} cannot be achieved without knowing initial conditions. However, the $\xi$ dependence of $S_F(p;\xi)$ can be deduced independently of the propagator itself at any specific gauge. 
	In the next two subsections we expand on these properties.

	\subsection{Spectral representation of fermion propagator\label{ss:spectral_rep_SF}}
	To find the solution to  \eqref{eq:Dxi_SF}, it is useful to recall the spectral representation for the fermion propagator. Up to subtractions this rewrites the propagator functions in the complex momentum plane as integrals of the free-particle propagator weighted by real functions. In addition, it allows direct evaluation of the effective one-loop integral on the right-hand side of  \eqref{eq:Dxi_SF}. Introducing the spectral representation of the fermion propagator also allows us to handle all $p^2$ dependences in  \eqref{eq:Dxi_SF} separately from the propagator, providing a promising candidate to solve for the Green's function of  \eqref{eq:Dxi_SF}.
	
	The Dirac scalar and Dirac vector components of a massive fermion propagator requires two spectral functions. For $S_F(p)=\slashed{p} S_1(p^2)+S_2(p^2)\mathbb{I}$,
	\begin{equation}
	S_j(p^2;\xi)=\int_{m^2}^{+\infty}ds\,\dfrac{\rho_j(s;\xi)}{p^2-s+i\varepsilon},\label{eq:fermion_spectral_rep}
	\end{equation}
	where from here on $j=1,2$ and the dependence of fermion propagator on the gauge fixing parameter $\xi$ has been made explicit. Since the spectral representation of the fermion propagator is independent of gauge fixing, the spectral functions $\rho_j$ embody the gauge dependence. That the integrals defined by Eq.~\eqref{eq:fermion_spectral_rep} converge without the need for subtractions is assured by the renormalizability of QED in $d < 4$ dimensions. When $d$ approaches 4, this remains so once the fermion propagator functions are multiplied by the appropriate combination of renormalization factors $Z_2$ and $Z_m$. This is most readily achieved using dimensional
	regularization. 
	
	The existence of a spectral representation in the form of  \eqref{eq:fermion_spectral_rep} is closely related to the allowed analytic structures of the fermion propagator. In terms of the complex variable $p^2$, $s$ is used to represent it when $\mathrm{Re}\{p^2\}\geq m^2$. Then, the fermion propagator functions are defined by their singularities, namely poles and branch cuts in the complex momentum plane. A simple pole corresponds to a free-particle contribution to the propagator. While branch cuts arise from  quantum loop corrections in the region where real particles are produced. We assume that for QED, the only singularities are branch cuts in the timelike region and a finite number of poles, with the propagator function holomorphic everywhere else. 
	
	\begin{figure}
	\centering
	\includegraphics[width=0.5\linewidth]{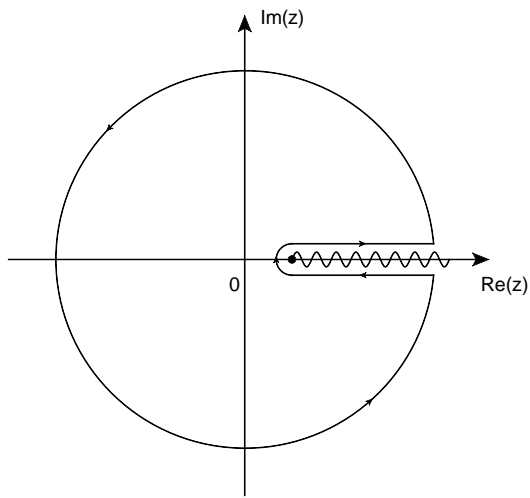}
	\caption{The illustration of analytic functions with branch cuts along the positive real axis. The contour can be used prove  \eqref{eq:fermion_spectral_rep} using Cauchy's integral formula when $z$ is replaced by $p^2$.}
	\label{fig:analytic_fz}
	\end{figure}
	It is apparent that simple poles located on the positive real axis of $p^2$ correspond to $\delta$-function terms in spectral functions. To see how a regular function (as opposed to $\delta$-functions being distributions) in $\rho(s)$ contributes to a branch cut along the positive real axis starting at the threshold $m^2$ through  \eqref{eq:fermion_spectral_rep}, consider subtracting pole structures from the propagator function first. Since the kernel function $1/(p^2-s+i\varepsilon)$ is holomorphic except for $p^2=s$, the remaining term in $S(p^2)$ is holomorphic in the complex momentum plane except for $p^2\geq m^2$. In addition, because $1/(p^2-s+i\varepsilon
	)=1/(\overline{p^2-s-i\varepsilon})$, close to the branch cut we have $S(p^2+i\varepsilon)=S^*(p^2-i\varepsilon)$. Therefore we know the imaginary part of $S(p^2)$ along the branch cut is antisymmetric going across the positive real axis from the first quadrant to the fourth quadrant. Taking the contour illustrated in Fig \ref{fig:analytic_fz}, when the contribution from the infinite radius arc vanishes, a direct application of Cauchy's integral formula shows that
	\begin{equation}
	\rho_j(s;\xi)=-\dfrac{1}{\pi}\mathrm{Im}\big\{S_j(s+i\varepsilon;\xi)\big\}.\label{eq:fermion_inv_spectral_rep}
	\end{equation}
	Apparently  \eqref{eq:fermion_inv_spectral_rep} applies when free-particle poles are present as well, since as is well known $\lim\limits_{\varepsilon\rightarrow 0}\varepsilon/(x^2+\varepsilon^2)=\pi\delta(x)$. When the propagators do not vanish for $p^2\rightarrow \infty$, substituting \eqref{eq:fermion_spectral_rep} into  \eqref{eq:S_j_subtraction} gives
	\begin{equation}
	\tilde{S}_j(p^2;\xi)=p^2\int_{m^2}^{+\infty}ds\dfrac{\rho_j(s;\xi)}{(p^2-s+i\varepsilon)s}, \label{eq:S_j_tilde}
	\end{equation}
	which is convergent if the $s\rightarrow +\infty$ limit of $\rho_j(s)$ is finite. When used to construct an ansatz for the fermion-photon vertex, the spectral functions can be solved from the fermion propagator SDE \cite{Delbourgo:1977jc}. The gauge dependence of these solutions has been explored by Delbourgo, Keck and Parker \cite{Delbourgo:1980vc}.

	\subsection{Various representations of LKFT}
	\begin{figure}
		\centering
		\begin{tikzpicture}
		\node[draw,circle] (A) at (90:3) {$\rho(s;\xi)$};
		\node[draw,circle] (B) at (210:3) {$D(p^2;\xi)$};
		\node[draw,circle] (C) at (330:3) {$P(x^2;\xi)$};
		\draw[-open triangle 45] (A.225) -- node[rotate=60,above] {$\int ds\dfrac{1}{p^2-s+i\varepsilon}$} (B.75);
		\draw[open triangle 45-] (A.255) -- node[rotate=60,below] {$-\dfrac{1}{\pi}\mathrm{Im}\{\}$} (B.45);
		\draw[-open triangle 45] (A.285) -- node[rotate=300,below] {...} (C.135);
		\draw[open triangle 45-] (A.315) -- node[rotate=300,above] {...} (C.105);
		\draw[-open triangle 45] (B.345) -- node[rotate=0,below] {$\mathcal{F}^{-1}$} (C.195);
		\draw[open triangle 45-] (B.15) -- node[rotate=0,above] {$\mathcal{F}$} (C.165);
		\end{tikzpicture}
		\caption{Scalar particles propagators in coordinate space $P(x^2;\xi)$, momentum space $D(p^2;\xi)$ and its spectral function $\rho(s;\xi)$ with bijective relations among them illustrated. The Fourier transform is bijective. For momentum space propagators with branch cuts and poles as their singularities, the spectral representation is also bijective. Consequently there must be a bijective relation between the coordinate space propagator and its spectral function.}
		\label{fig:LKFT_group}
		\end{figure}
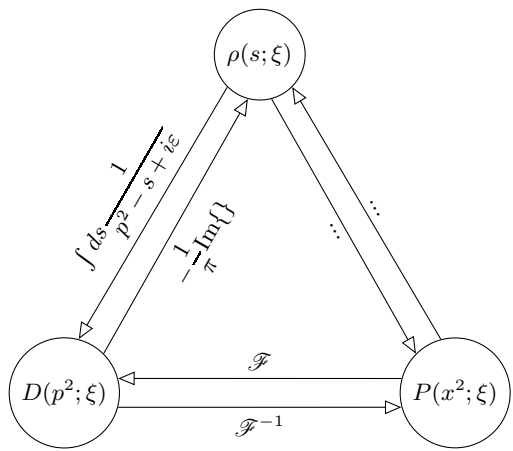
	Substituting the spectral representation of fermion propagator, \eqref{eq:fermion_spectral_rep}, into \eqref{eq:Dxi_SF} allows the effective one-loop integral to be evaluated explicitly. However, the spectral representation alone is not sufficient for us to solve for the dependence of fermion propagator on the gauge parameter $\xi$ from LKFT. 
	
	Observations about  \eqref{eq:LKFT_SF} will provide insight into a more useful mathematical aspect of LKFT for a gauge covariant  fermion propagator. Formally, \eqref{eq:LKFT_SF} states that the LKFT for the fermion propagator in coordinate space is simply a phase factor, which bears close resemblance to elements of a Lie group. 
	One can further verify that, when group multiplication is defined as function multiplication, this phase factor satisfies closure, associativity, and the existence of identity element and inverse elements. Therefore when considered as a linear transformation on coordinate space functions, the LKFT is indeed a group transform for coordinate space fermion propagators. 
	
	Fourier transforms are known to be {\it one-to-one} and {\it onto}. Since we have established in Subsection \ref{ss:spectral_rep_SF} that with certain assumptions about the analytic structure of the fermion propagator, the spectral representation is also {\it one-to-one} and {\it onto}. These correspondences, illustrated in Fig. \ref{fig:LKFT_group}, clearly indicate that, just as with the LKFT for coordinate space propagator, LKFT for momentum space propagators and for spectral functions should both be group transforms. In fact, the coordinate space representation, the momentum space representation and the spectral representation of LKFT are isomorphic representations of the same group. Additionally, since $\xi$ parameterizes the LKFT as a continuous group, the starting gauge of LKFT does not matter; only the difference in $\xi$ enters in calculation. Though the default initial gauge for LKFT can be conveniently chosen to be the Landau gauge,  for calculations with the  initial value of gauge parameter that is $\xi_0$, one simply replaces Landau gauge quantities by those at $\xi_0$ and replaces $\xi$ by $\xi-\xi_0$.
	
	Having established that the LKFT in its spectral representation is a group transformation, we can proceed to develop schemes for solving \eqref{eq:Dxi_SF}. As illustrated in Fig.~\ref{fig:LKFT_group}, the correspondence between the fermion propagator in momentum space and its spectral function is linear, as a consequence of which LKFT in spectral form is also required to be linear. However, instead of a simple phase factor, we expect the LKFT in its spectral form to involve more complicated linear operations. Therefore, without loss of generality, we can write
	\begin{equation}
	\rho_j(s;\xi)=\int ds'\, \mathcal{K}_j(s,s';\xi)\,\rho_j(s';0),\label{eq:LKFT_linearity_spectral_rep}
	\end{equation}
	where distributions $\mathcal{K}_j(s,s';\xi)$ work as the Green's function for \eqref{eq:Dxi_SF}. They represent linear operations that encode $\xi$ dependences of $\rho_j(s;\xi)$ to be determined by the LKFT, and so respect all group properties. Explicitly, denote $\mathbf{K}$ the set of distribution $\mathcal{K}(s,s';\xi)$, with group multiplication defined as integration over spectral variables. To verify that $\mathbf{K}$ is indeed a group, for any $\mathcal{K}(s,s';\xi)\in \mathbf{K}$ the following properties have to be satisfied:
	\begin{enumerate}
	\item \underline{Closure} $\int ds'\mathcal{K}(s,s';\xi)\mathcal{K}(s',s'';\xi')$ is also an element of $\mathbf{K}$;
	\item \underline{Associativity} 
\begin{align*}\int ds'\mathcal{K}(s,s';\xi)\int ds''&\mathcal{K}(s',s'';\xi')\mathcal{K}(s'',s''';\xi'')\\
&=\int ds''\left[\int ds'\mathcal{K}(s,s';\xi)\mathcal{K}(s',s'';\xi')\right]\mathcal{K}(s'',s''';\xi'');
\end{align*}
	\item \underline{Identity Element} $\exists~\mathcal{K}_I(s,s') \in \mathbf{K}$ such that \[\int ds'\mathcal{K}_I(s,s')\mathcal{K}(s',s'';\xi)=\int ds'\mathcal{K}(s,s';\xi)\mathcal{K}_I(s',s'')=\mathcal{K}(s,s'';\xi);\]
	\item \underline{Inverse Element} $\exists~\mathcal{K}_{inv}(s,s';\xi)$ such that \[\int ds'\,\mathcal{K}_{inv}(s,s';\xi)\,\mathcal{K}(s',s'';\xi)\;=\;\int ds'\,\mathcal{K}(s,s';\xi)\,\mathcal{K}_{inv}(s',s'';\xi)\;=\;\mathcal{K}_I(s,s'').\]
	\end{enumerate}	
Substituting (\ref{eq:fermion_spectral_rep}~,\ref{eq:LKFT_linearity_spectral_rep}) into \eqref{eq:Dxi_SF} gives
	\begin{equation}
	\dfrac{\partial}{\partial\xi}\int ds\,\dfrac{\mathcal{K}_j(s,s';\xi)}{p^2-s+i\epsilon}\;=\;-\dfrac{\alpha}{4\pi}\int ds\,\dfrac{\Xi_j(p^2,s)}{p^2-s+i\epsilon}\,\mathcal{K}_j(s,s';\xi),\label{eq:LKFT_k12}
	\end{equation}
	where the $\Xi_j(p^2,s)$ are determined by the effective one-loop integral, which can be evaluated using Feynman parameterization for combining denominators, together with dimensional regularization. Apparently from \eqref{eq:LKFT_linearity_spectral_rep}, the initial condition for distributions $\mathcal{K}_j$ is $\mathcal{K}_j(s,s';0)=\delta(s-s')$. In the remaining part of this subsection, two methods for solving \eqref{eq:LKFT_k12} will be presented.
	\paragraph{Method 1: analogue to first-order ordinary differential equations} Operations with respect to $\xi$ in \eqref{eq:LKFT_k12} are only present on the left-hand side, which resemble homogeneous first-order ordinary differential equations. In order to solve for $\mathcal{K}_j(s,s';\xi)$, consider the original definition of partial derivative:
	\begin{equation}
	\dfrac{\partial}{\partial\xi}\mathcal{K}(s,s';\xi)\equiv \lim\limits_{\Delta\rightarrow 0}\dfrac{\mathcal{K}(s,s';\xi+\Delta)-\mathcal{K}(s,s';\xi)}{\Delta}.
	\end{equation}
	Next, applying \eqref{eq:LKFT_linearity_spectral_rep} many times gives
	\begin{align*}
	\rho(s,s';\xi+\Delta)& =\int ds''\mathcal{K}(s,s'';\Delta)\rho(s'',s';\xi)\\
	& =\int ds''\int ds'''\mathcal{K}(s,s'';\Delta)\mathcal{K}(s'',s''';\xi)\rho(s'',s';0).
	\end{align*}

	Since the LKFT is independent of initial conditions,
	\begin{equation}
	\mathcal{K}(s,s';\xi+\Delta)=\int ds''\mathcal{K}(s,s'';\Delta)\mathcal{K}(s'',s';\xi).\label{eq:k_sum}
	\end{equation}
	Equation~\eqref{eq:k_sum} should not come as a surprise given that the LKFT for the fermion propagator in a spectral representation is isomorphic to coordinate space LKFT. \eqref{eq:LKFT_k12} then becomes
	\begin{align}
	\dfrac{\partial}{\partial\xi}\int ds\dfrac{\mathcal{K}(s,s';\xi)}{p^2-s+i\epsilon}& =\lim\limits_{\Delta\rightarrow 0}\int ds\int ds''\dfrac{\mathcal{K}(s,s'';\Delta)-\delta(s-s'')}{(p^2-s+i\epsilon)\Delta}\mathcal{K}(s'',s';\xi)\nonumber\\[1.5mm]
	& =-\dfrac{\alpha}{4\pi}\int ds\dfrac{\Xi(p^2,s)}{p^2-s+i\epsilon}\mathcal{K}(s,s';\xi).\label{eq:dxi_Delta}
	\end{align}
	Taking the limit $\xi\rightarrow 0$ where $\mathcal{K}(s,s';\xi)$ becomes a  delta function simplifies \eqref{eq:dxi_Delta} into
	\begin{equation}
	\lim\limits_{\Delta\rightarrow 0}\dfrac{1}{\Delta}\int ds\dfrac{\mathcal{K}(s,s';\Delta)-\delta(s-s')}{p^2-s+i\epsilon}=-\dfrac{\alpha}{4\pi}\dfrac{\Xi(p^2,s')}{p^2-s'+i\epsilon}.\label{eq:dxi0_Delta}
	\end{equation}
	Equation~\eqref{eq:dxi0_Delta} specifies how the distribution $\mathcal{K}(s,s';\xi)$ departs from its initial form (a delta function) with infinitesimal $\xi$. Solving \eqref{eq:dxi0_Delta} is sufficient to obtain $\mathcal{K}(s,s';\xi)$ with finite $\xi$, which, in principle, can be written as an infinite number of steps of distribution multiplication. Explicitly, this procedure is
	\begin{equation}
	\mathcal{K}(s,s';\xi)=\lim\limits_{N\rightarrow +\infty}\left[\prod_{n=0}^{N-1}\int ds_{n+1}~\mathcal{K}\left(s_n,s_{n+1};\dfrac{\xi}{N}\right) \right]\mathcal{K}\left(s_{N},s';0\right),\label{eq:k_Nsteps}
	\end{equation}
	with $s_0=s$.
	Formally \eqref{eq:k_Nsteps} gives distributions $\mathcal{K}_j(s,s';\xi)$ with finite $\xi$, solving LKFT for $\rho_j(s;\xi)$. In practice one may prefer a closed form for the $\mathcal{K}_j$ with group multiplications in a minimal number of steps. Realizing \eqref{eq:k_Nsteps} is the analogue of 
	\[\lim\limits_{N\rightarrow +\infty}\left(1+\dfrac{x}{N}\right)^N=e^x,\]
	and \eqref{eq:LKFT_k12} is very  similar to $\dfrac{d}{dx}f(x)=af(x)$, we can assume the following form for $\mathcal{K}(s,s';\xi)$,
	\begin{equation}
	\mathcal{K}_j=\exp\left(-\dfrac{\alpha\xi}{4\pi}\Phi_j\right),\label{eq:k_exponential}
	\end{equation}
	where distributions $\Phi_j$ are independent of $\xi$. The exponential of a distribution is given by definition
	\begin{equation}
	\exp\big\{\lambda\Phi\big\}=\sum_{n=0}^{+\infty}\dfrac{\lambda^n}{n!}\Phi^n=\delta(s-s')+\lambda\Phi+\dfrac{\lambda^2}{2!}\Phi^2+\dots,\label{eq:dist_exp}
	\end{equation}
	with distribution exponentiation defined as
	\begin{equation}
	\Phi^0(s,s')=\delta(s-s')\quad \mathrm{and}\quad \Phi^n(s,s')=\int ds''\Phi(s,s'')\Phi^{n-1}(s'',s')\quad (n\geq 1).\label{eq:dist_exponentiation}
	\end{equation}
	One can check that $\mathcal{K}_j$ given by \eqref{eq:k_exponential} satisfy \eqref{eq:LKFT_k12} with initial conditions ${\mathcal{K}_j(s,s';0)=\delta(s-s')}$ given distributions $\Phi_j$ satisfy their own identities. To verify the exponential of distributions indeed solves \eqref{eq:LKFT_k12} and find the identities $\Phi_j$ have to satisfy, let us start with
	\[\dfrac{\partial}{\partial\xi}\mathcal{K}=\dfrac{\partial}{\partial\xi}\exp\bigg\{-\dfrac{\alpha\xi}{4\pi}\Phi\bigg\}=-\dfrac{\alpha}{4\pi}\Phi\exp\bigg\{-\dfrac{\alpha\xi}{4\pi}\Phi\bigg\}=-\dfrac{\alpha}{4\pi}\Phi \mathcal{K},\]
	then using \eqref{eq:k_exponential}, the left hand side of \eqref{eq:LKFT_k12} can be written as
	\begin{equation}
	\dfrac{\partial}{\partial\xi}\dfrac{1}{p^2-s+i\epsilon}\mathcal{K}=-\dfrac{\alpha}{4\pi}\dfrac{\Phi}{p^2-s+i\epsilon}\mathcal{K}.
	\end{equation}
	Comparing with its right hand side, one obtains after restoring the integration variables,
	\begin{equation}
	\int ds\,ds'\,\dfrac{\Phi(s,s')}{p^2-s+i\epsilon}\,\mathcal{K}(s',s'';\xi)=\int ds'\,\dfrac{\Xi(p^2,s')}{p^2-s'+i\epsilon}\,\mathcal{K}(s',s'';\xi),
	\end{equation}
	which indicates (or by multiplying $\mathcal{K}(s'',s''';-\xi)$ to the right)
	\begin{equation}
	\int ds\,\dfrac{\Phi_j(s,s')}{p^2-s+i\epsilon}\;=\;\dfrac{\Xi_j(p^2,s')}{p^2-s'+i\epsilon}.\label{eq:Phi_Xi}
	\end{equation}
	Therefore \eqref{eq:LKFT_k12} is solved by \eqref{eq:k_exponential} given $\Phi_j$ satisfy \eqref{eq:Phi_Xi}. One can easily identify that distributions $\Phi_j$ are the generators for continuous groups defined by $\mathcal{K}_j$.
	\paragraph{Method 2: differential equations solved by multiplying inverse elements}
	From the group property of $\mathbf{K}=\{\mathcal{K}(s,s';\xi) \}$, the inverse element of $\mathcal{K}(s,s';\xi)$ is $\mathcal{K}(s,s';-\xi)$. This can be seen most easily from the isomorphic representation of LKFT in coordinate space. Or, in the language of distribution multiplication,
	\[\mathcal{K}^{-1}(s,s';\xi)=\mathcal{K}(s,s';-\xi). \]
	Multiplying this inverse element to the right of differential equation \eqref{eq:LKFT_k12} gives
	\begin{align}
&	\int ds\int ds'\dfrac{1}{p^2-s+i\epsilon}\left[\dfrac{\partial}{\partial\xi}\mathcal{K}(s,s';\xi)\right]\mathcal{K}(s',s'';-\xi)\nonumber\\
&\hspace{3.5cm}=-\dfrac{\alpha}{4\pi}\int ds\int ds'\dfrac{\Xi(p^2,s)}{p^2-s+i\epsilon}\mathcal{K}(s,s';\xi)\mathcal{K}(s',s'';-\xi),
	\end{align}
	or equivalently,
	\begin{equation}
	\int ds\dfrac{1}{p^2-s+i\epsilon}\dfrac{\partial}{\partial\xi}\ln~\mathcal{K}(s,s'';\xi)=-\dfrac{\alpha}{4\pi}\int ds\dfrac{\Xi(p^2,s'')}{p^2-s''+i\epsilon},\label{eq:log_k}
	\end{equation}
	where the logarithm of distribution $\mathcal{K}(s,s'';\xi)$ is taken in distributional sense, hence when spectral variables are omitted
	\begin{equation}
	\ln (\mathcal{K})=(\mathcal{K}-\delta)-\dfrac{1}{2}(\mathcal{K}-\delta)^2+\dfrac{1}{3}(\mathcal{K}-\delta)^3+\dots=\sum_{n=1}^{+\infty}\dfrac{(-1)^{n-1}}{n}(\mathcal{K}-\delta)^n,
	\end{equation}
	with distribution exponentiations defined by \eqref{eq:dist_exponentiation}. To see that $(\partial_\xi~\mathcal{K})\mathcal{K}^{-1}$ is indeed $\partial_\xi\ln~\mathcal{K}$, denote $u=\ln \mathcal{K} \quad \mathcal{K}=e^u$. Then
	$\partial_\xi \mathcal{K}=(\partial_\xi u)e^u=(\partial_\xi u)\mathcal{K}$. Therefore $\partial_\xi \ln \mathcal{K}=\partial_\xi u=(\partial_\xi \mathcal{K})\mathcal{K}^{-1}$.
	
	After clarifying the meaning of distribution logarithm, the null space of the spectral representation is supposed to be empty for the function space defined as the set of functions with analytic structures discussed in Subsection \ref{ss:spectral_rep_SF}. Therefore \eqref{eq:log_k} indicates that when
	\begin{equation}
	\partial_\xi \ln~\mathcal{K}(s,s'';\xi)=-\dfrac{\alpha}{4\pi}\Phi(s,s''),
	\end{equation}
	with distribution $\Phi$ satisfying \eqref{eq:Phi_Xi}, \eqref{eq:Dxi_SF} is solved by \eqref{eq:k_exponential}.
	
	Either through group operations or analogy with ordinary differential equations, we have formally found the Green's function specifying the $\xi$ dependence of the fermion propagator spectral functions. Because there are no dimension-odd operators in the LKFT for the fermion propagator, the Dirac vector and Dirac scalar components do not mix. The representation of linear operations by integrating distributions with spectral functions closely resembles  matrices multiplying vectors as linear transforms.
	
\section{LKFT in spectral representation with arbitrary numbers of dimensions\label{ss:LKFT_spectral_rep}}
	Before applying these solutions to the LKFT in the form of \eqref{eq:k_exponential} to calculate the $\xi$ dependence of the fermion propagator, we need to determine the  distributions $\Phi_j$ from \eqref{eq:Phi_Xi}. To do so requires explicit expressions for the functions $\Xi_j(p^2,s)$. 
	
	We use standard perturbative techniques including the Feynman method for combining denominators and dimensional regularization. Then substituting the spectral representation of the fermion propagator \eqref{eq:fermion_spectral_rep} into the LKFT for the momentum space fermion propagator, \eqref{eq:Dxi_SF}, and comparing the resulting equation with the definition of $\Xi_j(p^2,s)$ in \eqref{eq:LKFT_k12} gives, 
	\begin{align}
	& \Xi_1(p^2,s)=\int_{0}^{1}dx~2x\left[1-\epsilon+\dfrac{\epsilon}{1-xz}\right]\dfrac{\Gamma(\epsilon)(4\pi\mu^2/s)^\epsilon}{[(1-x)(1-xz)]^\epsilon}\label{eq:def_Xi_1}\\
	& \Xi_2(p^2,s)=\int_{0}^{1}dx~2x\left[1-\epsilon+\dfrac{\epsilon}{2}\dfrac{z+1}{1-xz}\right]\dfrac{\Gamma(\epsilon)(4\pi\mu^2/s)^\epsilon}{[(1-x)(1-xz)]^\epsilon},\label{eq:def_Xi_2}
	\end{align}
	where $z=p^2/s$ and the number of spacetime dimensions $d=4 -2 \epsilon$. Meanwhile, the dimension of $e^2/(4\pi)$ is carried by $\mu$ such that the coupling constant $\alpha$ remains dimensionless.
	
	Results given in (\ref{eq:def_Xi_1},~\ref{eq:def_Xi_2}) characterize how the LKFT behaves in Minkowski space. Using the spectral representation there is no need to make a Wick rotation to perform the loop-type integral. This eliminates any ambiguity of which loop momentum should be integrated in Euclidean space. We use dimensional regularization (required when close to four dimensions) in one of two ways. We can follow Feynman and integrate the time component of the loop momentum to infinity first. We then have spherical symmetry in the $(d-1)$ spatial dimensions and use dimensional regularization only on the space components. Of course, we could instead Wick rotate, assuming this is valid and picks up no new singularities. One then has spherical symmetry in $d$ dimensions and regularize {\it \`a la} 't Hooft and Veltman \cite{'tHooft:1972fi}. The results are the same with or without Wick rotation, as discussed in Appendix \ref{ss:loop_Minkowski}. 

Extensive use of definitions and properties of hypergeometric functions allows us to evaluate integrals over Feynman parameters in (\ref{eq:def_Xi_1},~\ref{eq:def_Xi_2}) explicitly, see Appendix \ref{ss:Xi_12_epsilon} for details. The results are 
	\begin{align}
	\dfrac{\Xi_1}{p^2-s}&=\dfrac{\Gamma(\epsilon)}{s}\left(\dfrac{4\pi\mu^2}{s}\right)^\epsilon\dfrac{-2}{(1-\epsilon)(2-\epsilon)}~_2F_1(\epsilon+1,3;3-\epsilon;z)\label{eq:Xi_1_reduced}\\
	\dfrac{\Xi_2}{p^2-s}&=\dfrac{\Gamma(\epsilon)}{s}\left(\dfrac{4\pi\mu^2}{s}\right)^\epsilon\dfrac{-1}{1-\epsilon}~_2F_1(\epsilon+1,2;2-\epsilon;z).\label{eq:Xi_2_reduced}
	\end{align}
	Hypergeometric functions occurring in (\ref{eq:Xi_1_reduced},~\ref{eq:Xi_2_reduced}) are understood to be given by the integral definition (15.3.1) in Abramowitz and Stegun \cite{abramowitz1964handbook}. For $\epsilon>0$, this integral definition is the analytic continuation of the series definition \eqref{eq:2F1_Taylor} with a branch cut\cite{abramowitz1964handbook} on the real axis of $z$ from $1$ to $+\infty$, a property one would expect for corrections to the fermion propagator. The scenario where $\epsilon<0$ is beyond the scope of this article.
	
	Explicit calculation shows that in three dimensions
	\begin{align}
	& \lim\limits_{\epsilon\rightarrow 1/2}\Xi_1(p^2,s)=2\pi\sqrt{\dfrac{\mu^2}{s}}\left[-\dfrac{z+1}{(z-1)z} +\dfrac{z-1}{z^{3/2}}\mathrm{arctanh}(\sqrt{z})\right]\\
	& \lim\limits_{\epsilon\rightarrow 1/2}\Xi_2(p^2,s)=-\dfrac{4\pi}{z-1}\sqrt{\dfrac{\mu^2}{s}} \quad ,
	\end{align}
	while for small $\epsilon$, \textit{i.e.} approaching four dimensions:
	\begin{align}
	& \Xi_1(p^2,s)=\dfrac{1}{\epsilon}-\gamma_E+\ln\left(\dfrac{4\pi \mu^2}{s}\right)+1-\dfrac{1}{z}-\left(1+\dfrac{1}{z^2}\right)\ln(1-z)+\mathcal{O}(\epsilon^1)\\
	& \Xi_2(p^2,s)=\dfrac{1}{\epsilon}-\gamma_E+\ln\left(\dfrac{4\pi \mu^2}{s}\right)-\left(1+\dfrac{1}{z}\right)\ln(1-z)+\mathcal{O}(\epsilon).
	\end{align}
	The $\epsilon\rightarrow 1/2$ limits can be calculated using identities listed in Appendix \ref{ss:identities_2F1_AS}. While the small $\epsilon$ expansions can be calculated according to Appendix \ref{ss:identities_2F1_epsilon}. Therefore $d=3$ and $4$ results have been recovered.
	
	With loop integrals $\Xi_j(p^2,s)$ calculated, the right-hand side of \eqref{eq:Phi_Xi} is elegantly represented by (\ref{eq:Xi_1_reduced},~\ref{eq:Xi_2_reduced}). The remaining task is to find the corresponding distributions $\Phi_j$ that solve \eqref{eq:Phi_Xi}. Since the distributions $\Phi_j$ are only allowed to be linear operators on the spectral variable $s$, solving \eqref{eq:Phi_Xi} is equivalent to generating convoluted $p^2$ dependences embedded in hypergeometric functions from that of a free-particle propagator. For $\epsilon>0$, the behavior of functions $\Xi_j/(p^2-s+i\varepsilon)$ in the limit $p^2\rightarrow s$ is more singular than the free-particle propagator. In fact, this singularity behaves as 
	\begin{equation}
	\lim\limits_{p^2\rightarrow s}\dfrac{\Xi_j(p^2,s)}{p^2-s+i\varepsilon}=\Gamma(\epsilon)\left(\dfrac{4\pi\mu^2}{s}\right)^\epsilon \dfrac{4^\epsilon}{\sqrt{\pi}}\Gamma(1-\epsilon)\Gamma(1/2+\epsilon)\left(1-\dfrac{p^2}{s}-i\varepsilon\right)^{-1-2\epsilon},\label{eq:LKFT_sgl_Sigma}
	\end{equation}
	based on (15.3.6) of Abramowitz and Stegun \cite{abramowitz1964handbook}, also listed as \eqref{eq:2F1_singular_z1} of this article. Therefore one can expect distributions $\Phi_j$ to be more singular than $\delta$-functions.
	\subsection{Exponent-preserving operations}
	Our task is to find out how to generate $p^2$ dependences in hypergeometric functions given by (\ref{eq:Xi_1_reduced},~\ref{eq:Xi_2_reduced}) from the free-particle propagator with only linear operations on the spectral variable $s$. It appears that the variable $z=p^2/s$ is more convenient than the spectral variable $s$ itself. In the process of finding the distributions $\Phi_j$, multiplication by $s$ can be regarded as a trivial linear operation. Therefore we are allowed to apply it as needed to make the remaining operations transparent. Meanwhile, having decided to work with the variable $z$ rather than the dispersive variable $s$, we are obliged to ensure that the net effect of operations on $z$ does result in any operation on $p^2$. 
	
	Starting with the observation that the $p^2$ dependence of a free-particle propagator can be represented by 
	\[\dfrac{-s}{p^2-s}= \dfrac{1}{1-z}=~_2F_1(1,b;b;z) \; ,\]
	for any $b$.
	The factor $-s$ does not matter in this scenario because it is merely a multiplication factor. In addition, for any linear operation on the variable $z$, as long as the net effect does not act as multiplication by the variable $z$ or $z^\lambda$, such transforms can be written in terms of spectral variable $s$ independently of $p^2$. 
	
	To quantify this criterion, define the exponent $\lambda$ for linear transforms on the variable $z$. Starting with a simple multiplication factor $z^\lambda$, this has an exponent  $\lambda$, because it raises the index by $\lambda$ for every term in the series expansion of a function of $z$. 
	Thus the operation $z^md^{n}/dz^{n}$ has an exponent $\lambda=m-n$. 
	
	An identity involving transformations on the variable $z$  is called \textit{exponent-preserving} if the exponent of the transform on the left-hand side is identical to that on the right-hand side. For example, (15.2.2) in Abramowitz and Stegun \cite{abramowitz1964handbook} (also listed in Appendix \ref{ss:identities_2F1}) is exponent-preserving for transformations on variable $z$ because exponents on both sides are identical. An identity that does not preserve exponents is called \textit{exponent-violating}. Exponent-violating transforms on variable $z$ cannot be translated into operations on the spectral variable $s$ only (not involving $p^2$).
	
	Next we need to determine the exponent-preserving linear transforms that generate any hypergeometric function $~_2F_1(a,b;c;z)$ from $~_2F_1(1,b;b;z)=1/(1-z)$. To accomplish this, one immediately thinks of Gauss' relations for contiguous functions. However, they only relate hypergeometric functions with integer differences of parameters $a,b$ and $c$. Meanwhile, not all of them are exponent-preserving. Another category of candidates is the differential relation for hypergeometric functions. These relations (15.2.3,~15.2.4) in Abramowitz and Stegun \cite{abramowitz1964handbook} are promising since they are exponent-preserving. However (15.2.3, 15.2.4) in Abramowitz and Stegun \cite{abramowitz1964handbook} can not be applied without generalization because they, similar to relations for contiguous functions, only raise or lower parameters $a,b$ or $c$ by integers.
	\subsection{Fractional calculus}
	To be able to solve LKFT in arbitrary dimensions, we need to overcome the limitation that (15.2.3, 15.2.4) in Abramowitz and Stegun \cite{abramowitz1964handbook} only work for integer differences in parameters for hypergeometric functions. Consequently we consider generalizing these to fractional orders of derivatives. We need to find out a version of fractional derivatives that applies to these differential relations for $~_2F_1(a,b;c;z)$. To do this, it is natural to consider the Riemann-Liouville definition of fractional calculus \cite{Riemann:fractional}:
	\begin{equation}
	I^\alpha f(z)=\dfrac{1}{\Gamma(\alpha)}\int_{\zeta}^{z}dz'(z-z')^{\alpha-1}f(z').\label{eq:def_Riemann_Liouville_I}
	\end{equation}
	For $\alpha>0$, the Riemann-Liouville fractional derivative is defined as
	\begin{equation}
	D^\alpha f(z)=\left(\dfrac{d}{dz}\right)^{\lceil \alpha \rceil}I^{\lceil\alpha\rceil-\alpha}f(z),
	\end{equation}
	where $\lceil \alpha \rceil$ is the smallest integer larger than $\alpha$, {\it i.e.} the ceiling function. Specifically for $\alpha \in (0,1)$, $\lceil \alpha \rceil=1$ and
	\begin{equation}
	D^\alpha f(z)=\dfrac{1}{\Gamma(1-\alpha)}\dfrac{d}{dz}\int_{\zeta}^{z}dz'(z-z')^{-\alpha}f(z').\label{eq:def_Riemann_Liouville_D}
	\end{equation}
	The lower limit $\zeta$ should be selected to reproduce (15.2.3, 15.2.4) in Abramowitz and Stegun \cite{abramowitz1964handbook} if the Riemann-Liouville formulation of fractional calculus is the expected version of fractional calculus that successfully generalizes them.
	
	To make an informed selection of $\zeta$, consider \eqref{eq:2F1_Taylor}, the Taylor series expansion of hypergeometric functions. For $\alpha\in (0,1)$,
	\begin{equation}
	D^\alpha z^\beta=\dfrac{1}{\Gamma(1-\alpha)}\dfrac{d}{dz}\int_{\zeta}^{z}dz'(z-z')^{-\alpha}(z')^\beta.
	\end{equation}
	Since mixing among terms of the Taylor expansion after derivative operations is undesirable, we choose $\zeta=0$ and obtain
	\begin{equation}
	D^\alpha z^\beta=\dfrac{1}{\Gamma(1-\alpha)}\dfrac{d}{dz} \dfrac{\Gamma(1-\alpha)\Gamma(1+\beta)}{\Gamma(2-\alpha+\beta)}z^{1-\alpha+\beta}=\dfrac{\Gamma(1+\beta)}{\Gamma(1-\alpha+\beta)}z^{-\alpha+\beta},
	\end{equation}
	which applies when $\alpha<1,~\beta>-1$ and $z>0$. Since directly from the definition of Pochhammer symbol $(1-\alpha+\beta)_\alpha=\Gamma(1+\beta)/\Gamma(1-\alpha+\beta)$, we have
	\begin{equation}
	D^\alpha z^\beta =(1-\alpha+\beta)_\alpha z^{-\alpha+\beta}.\label{eq:fractional_D_alpha}
	\end{equation}
	Notice that derivatives generalized this way to fractional orders also agree with integer order derivatives when $\alpha$ in \eqref{eq:fractional_D_alpha} is an integer. This will be taken as the default definition of fractional calculus in this article.
	
	Showing (15.2.3, 15.2.4) in Abramowitz and Stegun \cite{abramowitz1964handbook} apply to fractional orders is straightforward starting with the application of Taylor series expansions of hypergeometric functions. Explicitly, 
	\[D^\alpha z^{a+\alpha-1}~_2F_1(a,b;c;z)=D^\alpha \sum_{n=0}^{+\infty}\dfrac{(a)_n(b)_n}{(c)_n n!}z^{n+a+\alpha-1}=\sum_{n=0}^{+\infty}\dfrac{(b)_n}{(c)_n n!}(a)_n(n+a)_\alpha z^{n+a-1}\]
	Since
	\[(a)_n(n+a)_\alpha=\dfrac{\Gamma(a+n)\Gamma(a+n+a)}{\Gamma(a)\Gamma(n+a)}=\dfrac{\Gamma(a+\alpha)\Gamma(a+n+\alpha)}{\Gamma(a)\Gamma(a+\alpha)}=(a)_\alpha(a+\alpha)_n, \]
	we have
	\begin{equation}
	D^\alpha z^{a+\alpha-1}~_2F_1(a,b;c;z)=\sum_{n=0}^{+\infty}(a)_\alpha z^{a-1}\dfrac{(a+\alpha)_n(b)_n}{(c)_n n!}z^n=(a)_\alpha z^{a-1}~_2F_1(a+\alpha,b;c;z).\label{eq:Da2F1}
	\end{equation}
	Therefore (15.2.3) in Abramowitz and Stegun \cite{abramowitz1964handbook} has been generalized to accommodate fractional orders of derivatives. Meanwhile, \eqref{eq:Da2F1} is exponent-preserving as one would expect. Therefore it is the generalization of (15.2.3) we are seeking. Similar steps can be used to prove that (15.2.4) in Abramowitz and Stegun \cite{abramowitz1964handbook} generalizes to fractional orders using our definition of fractional calculus as well;
	\begin{align}
	& \quad D^\alpha z^{c-1}~_2F_1(a,b;c;z)\nonumber\\
	& =D^\alpha\sum_{n=0}^{+\infty}\dfrac{(a)_n(b)_n}{(c)_n n!}z^{n+c-1}=\sum_{n=0}^{+\infty}\dfrac{(a)_n(b)_n}{(c)_nn!}(n+c-\alpha)_\alpha z^{n+c-1-\alpha}\nonumber\\
	& =\sum_{n=0}^{+\infty}\dfrac{(a)_n(b)_n}{n!}\dfrac{\Gamma(c)}{\Gamma(c+n)}\dfrac{\Gamma(n+c)}{\Gamma(n+c-\alpha)}z^{n+c-1-\alpha}\nonumber\\
	& =\sum_{n=0}^{+\infty}\dfrac{(a)_n(b)_n}{n!}\dfrac{\Gamma(c)}{\Gamma(c-\alpha)}\dfrac{\Gamma(c-\alpha)}{\Gamma(n+c-\alpha)}z^n z^{c-\alpha-1}\nonumber\\[1.5mm]
	& =(c-\alpha)_\alpha z^{c-\alpha-1}~_2F_1(a,b;c-\alpha;z),\label{eq:Dc2Fa}
	\end{align}
	because the hypergeometric function $~_2F_1(a,b;c;z)$ is symmetric in parameters $a$ and $b$. Equipped with (\ref{eq:Da2F1},~\ref{eq:Dc2Fa}), any hypergeometric function $~_2F_1(a,b;c;z)$ can be linearly generated from the free-particle propagator with only a finite (up to two) steps of exponent-preserving linear operations. Explicitly, to generate the $z$ dependences of $~_2F_1(\epsilon+1,n;n-\epsilon;z)$ from a free particle propagator, consider the following linear operations
	\begin{align}
	& D^\epsilon z^\epsilon ~_2F_1(1,n;n;z)=(1)_\epsilon~_2F_1(1+\epsilon,n;n;z)\label{eq:Da_epsilon}\\[1mm]
	& D^\epsilon z^{n-1}~_2F_1(1+\epsilon,n;n;z)=(n-\epsilon)_\epsilon
	z^{n-\epsilon-1}~_2F_1(1+\epsilon,n;n-\epsilon;z)\label{eq:Dc_epsilon}.
	\end{align}
	Therefore
	\begin{equation}
	~_2F_1(1+\epsilon,n;n-\epsilon;z)=\dfrac{\Gamma(n-\epsilon)}{\Gamma(n)\Gamma(1+\epsilon)}z^{\epsilon+1-n}D^\epsilon z^{n-1}D^\epsilon z^\epsilon~_2F_1(1,n;n;z),\label{eq:linear_trans_LKFT_epsilon}
	\end{equation}
	which after setting $n=2,3$ for $j=2,1$ respectively, recovers the exponent-preserving linear transforms required to generate the hypergeometric functions in $\Xi_j(p^2,s)/(p^2-s+i\varepsilon)$ from the free-particle propagator. To see explicitly how such a linear transform on $z$ can be written as that on $s$ not involving $p^2$, refer to Appendix \ref{ss:Example_epn_preserving} for an example. In addition, the exact order of component transforms given by (\ref{eq:Da_epsilon},~\ref{eq:Dc_epsilon}) should not matter because of the commutation relations for the hypergeometric function $~_2F_1(1,n;n;z)$
	\begin{equation}
	[z^{\epsilon+1-n}D^\epsilon z^{n-1},D^\epsilon z^\epsilon]~_2F_1(1,n;n;z)=0.
	\end{equation}
	This commutation relation is true for $~_2F_1(1,n;n;z)$ because (\ref{eq:Da_epsilon},~\ref{eq:Dc_epsilon}) acts on parameters $a$ and $c$ of the hypergeometric function $~_2F_1(a,b;c;z)$ independently.

	\subsection{Operator Exponentials\label{ss:Exp_k}}
	The combination of exponent-preserving requirement and fractional calculus allows us to solve for distributions $\Phi_j$ from \eqref{eq:Phi_Xi}. The solution for the fermion propagator LKFT in spectral form is given by the exponential of distributions written formally as \eqref{eq:k_exponential}. Using the definition of the distribution exponential in \eqref{eq:dist_exp}, one can calculate $\mathcal{K}_j(s,s';\xi)$ to any order in $\xi$. However, such expansions only converge well for small $\alpha\xi$, and it is difficult to calculate at high orders. 
	
	The result for the distributions $\mathcal{K}_j(s,s';\xi)$ acting on an arbitrary function of spectral variables might be difficult to calculate. However, for massive fermion propagators, because their singularities do not occur before the mass threshold, Taylor expansions of such functions about $p^2=0$ always have finite radii of convergence. Therefore for the purpose of finding the gauge covariance condition for the fermion propagator, once we know how distributions $\mathcal{K}_j$ act on $z^\beta$, sufficiently with any $\beta\in \mathbf{Z}$, we know the distribution completely.
	
To start, let us consider the following identity:
	\begin{equation}
	\dfrac{1}{p^2-s+i\varepsilon}=-\dfrac{z}{p^2}~_2F_1(1,b;b;z)\, ,\label{eq:free_particle_prop_2F1}
	\end{equation} 
where recall $z \equiv p^2/s$.
	Because exponent-preserving operations on $z$ do not have any net effect on $p^2$, we are allowed to multiply by $p^2$ on both sides of \eqref{eq:Phi_Xi}, which then becomes
	\begin{equation}
	\int ds'\,\Phi\dfrac{z}{z-1+i\varepsilon}=\dfrac{p^2\,\Xi(p^2,s)}{p^2-s+i\varepsilon}.
	\end{equation}
	We define the dimensionless operator $\phi$ such that at the operator level $\int ds'\Phi=\phi$. Then
	\begin{equation}
	\phi\dfrac{z}{z-1+i\varepsilon}=\dfrac{p^2\,\Xi}{p^2-s+\varepsilon}.\label{eq:phiXi}
	\end{equation}
	Next, substituting \eqref{eq:free_particle_prop_2F1} into \eqref{eq:phiXi} and combining the result with (\ref{eq:Xi_1_reduced},\ref{eq:Xi_2_reduced}) and \eqref{eq:linear_trans_LKFT_epsilon} gives
	\begin{equation}
	-\phi_n z~_2F_1(1,n;n;z)=\Gamma(\epsilon)\left(\dfrac{4\pi\mu^2}{p^2}\right)^\epsilon \dfrac{-\Gamma(2-\epsilon)}{(1-\epsilon)\Gamma(1+\epsilon)}z^{2\epsilon+2-n}D^\epsilon z^{n-1}D^\epsilon z^{\epsilon-1}z~_2F_1(1,n;n;z),
	\end{equation}
	from which we have
	\begin{equation}
	\phi_n=\Gamma(\epsilon)\left(\dfrac{4\pi\mu^2}{p^2}\right)^\epsilon \dfrac{\Gamma(1-\epsilon)}{\Gamma(1+\epsilon)}z^{2\epsilon+2-n}D^\epsilon z^{n-1}D^\epsilon z^{\epsilon-1}.\label{eq:def_phi}
	\end{equation}
	The distributions $\phi_n$ in \eqref{eq:def_phi} correspond to $\Phi_j$ with $j=1,2$ when $n=3,2$ respectively.
	
	With the explicit form of $\Phi_j$ known as \eqref{eq:def_phi}, we can proceed to calculate their exponentials. For convenience, define the operational part of $\phi_n$ as 
	\begin{equation}
	\overline{\phi}_n\equiv z^{2\epsilon+2-n}D^\epsilon z^{n-1}D^\epsilon z^{\epsilon-1}
	\end{equation}
	The action of $\overline{\phi}_n$ on $z^\beta$ can be calculated directly;
	\begin{align}
	 \quad \overline{\phi}_nz^\beta &= z^{2\epsilon+2-n}D^\epsilon z^{n-1}D^\epsilon z^{\beta+\epsilon-1}
	\quad=\quad (\beta)_\epsilon z^{2\epsilon+2-n}D^\epsilon z^{n+\beta-2}\nonumber\\
	&=(\beta)_\epsilon (n+\beta-\epsilon-1)_\epsilon z^{\beta+\epsilon}\;
	\quad =\quad\dfrac{\Gamma(n+\beta-1)\Gamma(\beta+\epsilon)}{\Gamma(n+\beta-\epsilon-1)\Gamma(\beta)}z^\epsilon z^\beta.\label{eq:phibar_beta}
	\end{align}

	For the purpose of finding out how distributions $\mathcal{K}_j$ act on $z^\beta$, we need an explicit expression for $\overline{\phi}^m_nz^\beta$, which can be obtained by applying \eqref{eq:phibar_beta} recursively,
	\begin{align}
	\overline{\phi}^m z^\beta& =\dfrac{\Gamma(n+\beta-1)\Gamma(\beta+\epsilon)}{\Gamma(n+\beta-\epsilon-1)\Gamma(\beta)}\phi^{m-1}z^{\beta+\epsilon}\nonumber\\
	& =\dfrac{\Gamma(n+\beta-1)\Gamma(\beta+\epsilon)}{\Gamma(n+\beta-\epsilon-1)\Gamma(\beta)}\dfrac{\Gamma(n+\beta+\epsilon-1)\Gamma(\beta+2\epsilon)}{\Gamma(n+\beta-1)\Gamma(\beta+\epsilon)}\dots\times\nonumber\\
	&\quad\quad\quad\quad\quad\quad\quad\quad \dfrac{\Gamma{(n+\beta+(m-1)\epsilon-1)}\Gamma(\beta+m\epsilon)}{\Gamma(n+\beta+(m-2)\epsilon-1)\Gamma(\beta+(m-1)\epsilon)}z^{\beta+m\epsilon}\nonumber\\
	& =z^{\beta+m\epsilon}\prod_{k=1}^{m}\dfrac{\Gamma(n+\beta+(k-1)\epsilon-1)\Gamma(\beta+k\epsilon)}{\Gamma(n+\beta+(k-2)\epsilon-1)\Gamma(\beta+(k-1)\epsilon)}\nonumber\\
	& =\dfrac{\Gamma(n+\beta+(m-1)\epsilon-1)\Gamma(\beta+m\epsilon)}{\Gamma(n+\beta-\epsilon-1)\Gamma(\beta)}z^{\beta+m\epsilon}.
	\end{align}	
	Alternatively, the calculation is more transparent by substituting $u=z^\epsilon, \quad \lambda=\beta/\epsilon$.
	\begin{align}
	\overline{\phi}^m_nu^{\lambda}& =u^{\lambda+m}\prod_{k=1}^{m}\dfrac{\Gamma(n+(\lambda+k-1)\epsilon)\Gamma((\lambda+k)\epsilon)}{\Gamma(n+(\lambda+k-2)\epsilon-1)\Gamma((\lambda+k-1)\epsilon)}\nonumber\\
	& =\dfrac{\Gamma(n+(\lambda+m-1)\epsilon-1)\Gamma((\lambda+m)\epsilon)}{\Gamma(n+(\lambda-1)\epsilon-1)\Gamma(\lambda\epsilon)}u^{\lambda+m}.
	\end{align}	
	After defining 
	\begin{equation}
	\overline{\alpha}\equiv \dfrac{\alpha\xi}{4\pi}\dfrac{\Gamma(\epsilon)\Gamma(1-\epsilon)}{\Gamma(1+\epsilon)}\left(\dfrac{4\pi\mu^2}{p^2}\right)^\epsilon,\label{eq:def_alpha}
	\end{equation}
	we obtain
	\begin{align}
	\mathcal{K}_j z^\beta& =\exp\left(-\dfrac{\alpha\xi}{4\pi}\phi_n\right)z^\beta=\exp\left(-\overline{\alpha}\overline{\phi}_n \right)z^\beta=\sum_{m=0}^{+\infty}\dfrac{(-\overline{\alpha})^m}{m!}\overline{\phi}^m_nz^\beta\nonumber\\
	& =\sum_{m=0}^{+\infty}\dfrac{(-\overline{\alpha})^m}{m!}\dfrac{\Gamma(n+\beta+(m-1)\epsilon-1)\Gamma(\beta+m\epsilon)}{\Gamma(n+\beta-\epsilon-1)\Gamma(\beta)}z^{\beta+m\epsilon},\label{eq:kn_zbeta}
	\end{align}	
	with $n=3,~2$ for $j=1,~2$ respectively. Equation \eqref{eq:kn_zbeta} specifies how $\mathcal{K}_n$ transforms one function of the spectral variable $s$ into another.  Notice $\overline{\alpha}$ always combines with $z^\epsilon$ to produce a factor of $(\mu^2/s)^\epsilon$, rendering $\mathcal{K}_j$ exponent-preserving. Therefore the action of $\mathcal{K}_j$ on any function can now be calculated as long as this function can be written as a linear combination of $s^{-\beta}$. This can be best understood through Mellin transforms. Effectively \eqref{eq:kn_zbeta} tells us what the Mellin transform of $\mathcal{K}_j$ is. Since a Mellin transform disentangles multiplicative convolutions, the action of $\mathcal{K}_j$ on any function of spectral variable $s$ can be reconstructed though the inverse Mellin transform.
	
	Combining the spectral representation for the fermion propagator, \eqref{eq:fermion_spectral_rep}, with the LKFT as a linear transform on spectral functions \eqref{eq:LKFT_linearity_spectral_rep} produces
	\begin{equation}
	S_j(p^2;\xi)=\int ds\int ds'\,\dfrac{1}{p^2-s+i\varepsilon}\,\mathcal{K}_{j}(s,s';\xi)\,\rho_j(s';0).\label{eq:LKFT_linearity_momentum}
	\end{equation}
	Because the group multiplication of $\mathbf{K}$ is associative, it does not matter which spectral integral in \eqref{eq:LKFT_linearity_momentum} is evaluated first. Looking only at the $s'$ integral, once the spectral functions of a fermion propagator at one covariant gauge is known, their counterparts at other covariant gauge can be calculated, which explains the meaning of \eqref{eq:LKFT_linearity_spectral_rep}. 
	
	Alternatively when considering $\mathcal{K}_j$ acting on the free-particle propagator $1/(p^2-s+i\varepsilon)$, it transforms the free propagator into a function of both $p^2$ and $\xi$. Directly applying \eqref{eq:kn_zbeta} gives
	\begin{align}
	\mathcal{K}_j\dfrac{1}{p^2-s+i\varepsilon} & =-\dfrac{1}{p^2}\mathcal{K}_j\dfrac{z}{1-z}\nonumber\\
	&  =-\dfrac{1}{p^2}\sum_{\beta=1}^{+\infty}\sum_{m=0}^{+\infty}\dfrac{(-\overline{\alpha})^m}{m!}\dfrac{\Gamma(n+\beta+(m-1)\epsilon-1)\Gamma(\beta+m\epsilon)}{\Gamma(n+\beta-\epsilon-1)\Gamma(\beta)}z^{\beta+m\epsilon}.\label{eq:kn_free_prop}
	\end{align}
	Substituting \eqref{eq:kn_free_prop} into \eqref{eq:LKFT_linearity_momentum} then produces
	\begin{equation}
	S_j(p^2;\xi)=-\int ds \dfrac{1}{p^2}\sum_{\beta=1}^{+\infty}\sum_{m=0}^{+\infty}\dfrac{(-\overline{\alpha})^m}{m!}\dfrac{\Gamma(n+\beta+(m-1)\epsilon-1)\Gamma(\beta+m\epsilon)}{\Gamma(n+\beta-\epsilon-1)\Gamma(\beta)}z^{\beta+m\epsilon}\rho_j(s;0),\label{eq:kn_SF}
	\end{equation}
	where as always $z=p^2/s$. Because for a given $\epsilon$, the imaginary part of \eqref{eq:kn_SF} can be calculated, the result reveals to what linear operations $\mathcal{K}_j$ correspond.
	
	For a specific number of dimensions, the function defined by \eqref{eq:kn_free_prop} as a double series could potentially be simplified. Special cases of 3D and 4D will be discussed in the following two sections.
	\section{LKFT for fermion propagator in 3D\label{ss:LKFT_fermion_3D}}
	When $d=3$, $\epsilon=1/2$, the effective one-loop integral in \eqref{eq:Dxi_SF} is finite. Without the ambiguity caused by infinite renormalization, LKFT in 3D can be solved directly, serving as an example to test claims about the general properties of LKFT in Sections \ref{ss:LKFT_group} and \ref{ss:LKFT_spectral_rep}.
	
	Starting with \eqref{eq:Dxi_SF}, after evaluating the effective loop-integral using the Feynman parameterization method, we obtain
	\begin{align}
	& \dfrac{\partial}{\partial\xi}\int ds\dfrac{\rho_1(s;\xi)}{p^2-s+i\epsilon}=\alpha\mu\int ds\bigg\{\dfrac{\sqrt{s}}{(p^2-s)^2}-\dfrac{\sqrt{s}}{2p^2(p^2-s)}\nonumber\\
	& {\hspace{5cm}}-\dfrac{1}{2(p^2)^{3/2}}\mathrm{arctanh}(\sqrt{p^2/s}) \bigg\}\rho_1(s;\xi)\label{eq:LKFT_fermion_3D_1}\\[2mm]
	& \dfrac{\partial}{\partial\xi}\int ds\dfrac{\rho_2(s;\xi)}{p^2-s+i\epsilon}=\alpha\mu\int ds\dfrac{\sqrt{s}}{(p^2-s)^2}\rho_2(s;\xi).\label{eq:LKFT_fermion_3D_2}
	\end{align}
	Since \eqref{eq:LKFT_fermion_3D_2} appears much simpler than \eqref{eq:LKFT_fermion_3D_1}, let us consider its solution first.
	
	Utilizing \eqref{eq:LKFT_linearity_spectral_rep}, the dependence of $\rho_2(s;\xi)$ on the covariant gauge parameter $\xi$ can be written as $\rho_2(s;\xi)=\int ds' \mathcal{K}_2(s,s';\xi)\rho_2(s';0)$. Since to generate $(p^2-s)^{-2}$ from $(p^2-s)^{-1}$ linearly involves a first order derivative, the distribution $\mathcal{K}_2$ should be given by 
	\begin{equation}
	\mathcal{K}_2(s,s';\xi)=\delta\left(s-\left(\sqrt{s'}+\alpha\mu\xi/2\right)^2\right),\label{eq:test_dist_2_3D}
	\end{equation}
	which corresponds to operations that shift and rescale the spectral function $\rho_2$. It is straightforward to show that the distribution $\mathcal{K}_2(s,s';\xi)$ satisfies its differential equation required by \eqref{eq:LKFT_fermion_3D_2}. Meanwhile, it reduces to a simple $\delta$-function when $\xi=0$. Therefore \eqref{eq:test_dist_2_3D} indeed specifies how $\rho_2$ changes from one covariant gauge to another. Additionally, $\mathcal{K}_2$ given by \eqref{eq:test_dist_2_3D} satisfies group properties trivially. 
	
	While \eqref{eq:LKFT_fermion_3D_1} is more complicated than \eqref{eq:LKFT_fermion_3D_2}, Bashir and Raya~\cite{Bashir:2002sp} have solved the LKFT in coordinate space assuming that in the Landau gauge propagator is  free. They deduced using Fourier transforms, that under this assumption the Dirac vector part of fermion propagator at any covariant gauge is given by (13) of Ref.~8
	\begin{align}
	&\quad S_1(p^2;\xi =B(p_E;\xi)/p_E^2\nonumber\\
	& =\dfrac{-1}{p_E^2+(m+\alpha\mu\xi/2)^2}-\dfrac{\alpha\mu\xi}{2}\dfrac{m+\alpha\mu\xi/2}{p_E^2[p_E^2+(m+\alpha\mu\xi/2)^2]}+\dfrac{\alpha\mu\xi}{2p_E^3}\arctan\left(\dfrac{p_E}{m+\alpha\mu\xi/2} \right)\nonumber\\
	& =\dfrac{1}{(m+\alpha\mu\xi/2)^2}\dfrac{1}{x-1}-\dfrac{\alpha\mu\xi/2}{(m+\alpha\mu\xi/2)^3}\left(\dfrac{1}{x-1}-\dfrac{1}{x} \right)\nonumber\\
	& \quad -\dfrac{\alpha\mu\xi/2}{(m+\alpha\mu\xi/2)^3}\dfrac{1}{x\sqrt{-x}}\arctan(\sqrt{-x})\nonumber\\
	& =\dfrac{m}{\left(m+\dfrac{\alpha\mu\xi}{2}\right)^3}\dfrac{1}{x-1}-\dfrac{\dfrac{\alpha\mu\xi}{2}}{\left(m+\dfrac{\alpha\mu\xi}{2}\right)^3}\dfrac{1}{x}\left[\dfrac{1}{\sqrt{x}}\mathrm{arctanh}(\sqrt{x})-1 \right],\label{eq:S1_3D_Bashir}
	\end{align}
	where $\mathrm{arctanh}(u)=(1/2)\ln[(1+u)/(1-u)]$,  $p_E=\sqrt{-p^2}$ and ${x=p^2/(m+\alpha\mu\xi/2)^2}$. Meanwhile, since 
	\begin{align}
	& \dfrac{\mathrm{arctan}(\sqrt{-x})}{\sqrt{-x}}=\dfrac{\mathrm{arctanh}(\sqrt{x})}{\sqrt{x}},\\
	& -\dfrac{1}{\pi}\mathrm{Im}\Big\{\dfrac{1}{x-1+i\epsilon}\Big\}=\delta(x-1),\\
	& -\dfrac{1}{\pi}\mathrm{Im}\Bigg\{\dfrac{1}{x+i\epsilon}\left[\dfrac{\mathrm{arctanh}(\sqrt{x+i\epsilon})}{\sqrt{x+i\epsilon}}-1 \right] \Bigg\}=-\dfrac{\theta(x-1)}{2x^{3/2}},
	\end{align}
	we can take a shortcut of solving \eqref{eq:LKFT_fermion_3D_1} by finding out the spectral function of \eqref{eq:S1_3D_Bashir} as
	\begin{equation}
	-\dfrac{1}{\pi}\mathrm{Im}\{S_1 \}=\dfrac{m}{\left(m+\dfrac{\alpha\mu\xi}{2}\right)^3}\,\delta(x-1)+\dfrac{\dfrac{\alpha\mu\xi}{2}}{\left(m+\dfrac{\alpha\mu\xi}{2} \right)^3}\,\dfrac{\theta(x-1)}{2x^{3/2}},\label{eq:rho1_3D_Bashir}
	\end{equation}
	with $x=s/(m+\alpha\mu\xi/2)^2$. The $\delta$-function term in \eqref{eq:rho1_3D_Bashir} corresponds to the free-particle term in \eqref{eq:S1_3D_Bashir}. While the $\theta$-function term in \eqref{eq:rho1_3D_Bashir} comes from the inverse hyperbolic tangent function that generates a branch cut. 
	
	Since \eqref{eq:rho1_3D_Bashir} reduces to a $\delta$-function when $\xi=0$, it also satisfies the differential equation \eqref{eq:LKFT_fermion_3D_1}, since its Fourier transform satisfies the coordinate equivalent of \eqref{eq:LKFT_fermion_3D_1}. While \eqref{eq:LKFT_fermion_3D_1} specifies exactly what conditions the distribution $\mathcal{K}_1(s,s';\xi)$ has to meet, $\mathcal{K}_1$ is given by \eqref{eq:rho1_3D_Bashir} with the modification that $m\rightarrow \sqrt{s'}$. Explicitly,
	\begin{equation}
	\mathcal{K}_1(s,s';\xi)=\dfrac{\sqrt{s'}}{\sqrt{s'}+\dfrac{\alpha\mu\xi}{2}}\,\delta\left(s-\left(\sqrt{s'}+\dfrac{\alpha\mu\xi}{2}\right)^2\right)+\dfrac{\alpha\mu\xi}{4s^{3/2}}\,\theta\left(s-\left(\sqrt{s'}+\dfrac{\alpha\mu\xi}{2}\right)^2\right).\label{eq:test_dist_1_3D}
	\end{equation}
	We therefore obtain the LKFT for $\rho_1$ in 3D with $\rho_1(s;\xi)=\int ds'\mathcal{K}_1(s,s';\xi)\rho_1(s';0)$. The direct proof that \eqref{eq:test_dist_1_3D} is the distribution we are seeking is lengthy. The detailed calculation is given in Appendix \ref{ss:k1_3D_diff}.

	Compared with $\mathcal{K}_2$ given by \eqref{eq:test_dist_2_3D}, linear operations given by \eqref{eq:test_dist_1_3D} are more convoluted. The $\delta$-function term in \eqref{eq:test_dist_1_3D} corresponds to shift and rescale operations on $\rho_1$. The operation brought by the $\theta$-function term corresponds to a convolution with the spectral function $\rho_1$. It is trivial to show that $\mathcal{K}_2$ given by \eqref{eq:test_dist_2_3D} meets the group properties listed in Section \ref{ss:LKFT_group}. While for $\mathcal{K}_1$ given by \eqref{eq:test_dist_1_3D}, the associativity property is obvious. The identity element is found exactly when $\xi=0$. The closure property is proved in detail in Appendix \ref{ss:k1_3D_closure}. Once the closure property is satisfied, the inverse element of $\mathcal{K}_1(s,s';\xi)$ is just $\mathcal{K}_2(s,s';-\xi)$. Therefore we have shown that the sets of functions by $\mathcal{K}_j$ with multiplication defined for distributions  are indeed continuous groups with the gauge parameter $\xi$ working as the group parameter. Up until now, the LKFT for the fermion propagator in 3D has been obtained without using the general solution in the form of \eqref{eq:k_exponential}. 
	
	Since in the special scenario when $n=2$ and $\epsilon=1/2$, \eqref{eq:phibar_beta} simplifies to
	\begin{equation}
	\lim\limits_{\epsilon\rightarrow 1/2}\overline{\phi_2}z^\beta=\beta z^{\beta+1/2}=z^{3/2}\dfrac{d}{dz}z^\beta,
	\end{equation}
	where we have written the action of $\phi_2$ on $z^\beta$ as an operator independent of $\beta$. Then  $\phi_2$ given by \eqref{eq:def_phi} reduces to
	\begin{equation}
	\lim\limits_{\epsilon\rightarrow 1/2}\phi_2=\dfrac{4\pi\mu}{\sqrt{p^2}}z^{3/2}\dfrac{d}{dz}=-2\pi\mu\dfrac{d}{ds^{1/2}},
	\end{equation}
	which, when combined with \eqref{eq:k_exponential}, produces
	\begin{equation}
	\lim\limits_{\epsilon\rightarrow 1/2}\mathcal{K}_2=\exp\left(\dfrac{\alpha\xi\mu}{2}\dfrac{d}{ds^{1/2}}\right).\label{eq:k2_3D_exp}
	\end{equation}
	After identifying \eqref{eq:k2_3D_exp} as the shifting operator for functions of $\sqrt{s}$ by $\alpha\xi\mu/2$, the result agrees with \eqref{eq:test_dist_2_3D}. In principle, a similar calculation can be carried out for $\mathcal{K}_1$ in 3D as well. However, in practice, multiple operations are required to obtain the corresponding $\phi_1$, the calculation of whose exponential is nontrivial. 
		
	Alternatively, to verify that \eqref{eq:k_exponential} with $\Phi_j$ given by \eqref{eq:def_phi} solves LKFT in 3D, we only need to show that the imaginary part of \eqref{eq:kn_SF} corresponds to distributions $\mathcal{K}_j$ in \eqref{eq:test_dist_1_3D} and \eqref{eq:test_dist_2_3D} in the limit $\epsilon\rightarrow 1/2$. 
	
	In the case $n=2$ and $\epsilon=1/2$, which leads to $\overline{\alpha}=\alpha\xi\mu/\sqrt{p^2}$, because of the duplication formula (6.1.18) in Abramowitz and Stegun \cite{abramowitz1964handbook}, $\Gamma(2z)=(2\pi)^{-1/2}2^{2z-1/2}\Gamma(z)\Gamma(z+1/2)$, Gamma functions in \eqref{eq:def_phi} simplify to
	\begin{equation}
	\dfrac{\Gamma(n+\beta+(m-1)\epsilon-1)\Gamma(\beta+m\epsilon)}{m!\Gamma(n+\beta-\epsilon-1)\Gamma(\beta)}=\dfrac{\Gamma(2\beta+m)}{2^m\Gamma(2\beta)m!}.
	\end{equation}
	Next,
	\begin{equation}
	\sum_{m=0}^{+\infty}\dfrac{\Gamma(2\beta+m)}{2^m\Gamma(2\beta)m!}(-\overline{\alpha}\sqrt{z})^m=\left(1+\dfrac{\overline{\alpha}}{2}\sqrt{z} \right)^{-2\beta},
	\end{equation}
	and so we have
	\begin{align}
	\mathcal{K}_2\dfrac{1}{p^2-s+i\varepsilon}& =-\dfrac{1}{p^2}\sum_{\beta=1}^{+\infty}\left(1+\dfrac{\overline{\alpha}}{2}\sqrt{z} \right)^{-2\beta}z^\beta\nonumber\\
	& =\dfrac{-1}{p^2}\dfrac{z}{(1+\overline{\alpha}\sqrt{z})^2-z}=\dfrac{1}{p^2-(\sqrt{s}+\alpha\mu\xi/2)^2}.
	\end{align}
	Since when operating on the free-particle propagator produces identical results, $\mathcal{K}_2$ given by \eqref{eq:k_exponential} with $\Phi_2$ given by \eqref{eq:def_phi} agrees with $\mathcal{K}_2$ given by \eqref{eq:test_dist_2_3D}.
	
	Similarly for $\mathcal{K}_1$, when $n=3$ and $\epsilon=1/2$, the Gamma functions in \eqref{eq:def_phi} simplify by firstly noting
	\begin{align}
	& \quad \dfrac{\Gamma(n+\beta+(m-1)\epsilon-1)\Gamma(\beta+m\epsilon)}{m!\Gamma(n+\beta-\epsilon-1)\Gamma(\beta)}=\dfrac{\Gamma(\beta+m/2+3/2)\Gamma(\beta+m/2)}{\Gamma(\beta+3/2)\Gamma(\beta)m!}\nonumber\\
	& =\dfrac{\beta+m/2+1/2}{\beta+1/2}\dfrac{\Gamma(\beta+m/2+1/2)\Gamma(\beta+m/2)}{\Gamma(\beta+1/2)\Gamma(\beta)m!}=\left(1+\dfrac{m}{2\beta+1} \right)\dfrac{\Gamma(2\beta+m)}{2^m\Gamma(2\beta)m!}.
	\end{align}
	In addition since
	\begin{equation} \sum_{m=0}^{+\infty}\dfrac{m}{2\beta+1}\dfrac{\Gamma(2\beta+m)}{2^m\Gamma(2\beta)m!}(-\overline{\alpha}\sqrt{z})^m=\dfrac{-\beta}{1+2\beta}\overline{\alpha}\sqrt{z}\left(1+\dfrac{\overline{\alpha}}{2}\sqrt{z} \right)^{-1-2\beta}\nonumber\\
	\end{equation}
	and
	\begin{equation}
	\sum_{\beta=1}^{+\infty}\dfrac{-\beta}{1+2\beta}\overline{\alpha}\sqrt{z}\left(1+\dfrac{\overline{\alpha}}{2}\sqrt{z} \right)^{-1-2\beta}z^\beta=\dfrac{\overline{\alpha}}{2}\left[\dfrac{-\sqrt{z}\left( 1+\dfrac{\overline{\alpha}}{2}\sqrt{z}\right)}{\left(1+\dfrac{\overline{\alpha}}{2}\sqrt{z}\right)^2-z}+\mathrm{arctanh}\left(\dfrac{\sqrt{z}}{1+\overline{\alpha}\sqrt{z}/2} \right) \right],
	\end{equation}
	we have
	\begin{align}
	& \quad \mathcal{K}_1\dfrac{1}{p^2-s+i\varepsilon} =-\dfrac{1}{p^2}\sum_{\beta=1}^{+\infty}\left[\left(1+\dfrac{\overline{\alpha}}{2}\sqrt{z} \right)^{-2\beta}- \dfrac{\beta}{1+2\beta}\overline{\alpha}\sqrt{z}\left(1+\dfrac{\overline{\alpha}}{2}\sqrt{z} \right)^{-1-2\beta}\right]z^\beta\nonumber\\
	& =\dfrac{1}{p^2-(\sqrt{s}+\alpha\mu\xi/2)^2}-\dfrac{\alpha\xi\mu}{2p^2}\left[\dfrac{\sqrt{s}+\alpha\xi\mu/2}{p^2-(\sqrt{s}+\alpha\xi\mu/2)^2}+\dfrac{1}{\sqrt{p
			^2}}\mathrm{arctanh}\left(\dfrac{\sqrt{p^2}}{\sqrt{s}+\alpha\xi\mu/2} \right) \right],\label{eq:K_1_free_prop_4D}
	\end{align}
	which agrees with \eqref{eq:S1_3D_Bashir}. Consequently it also agrees with \eqref{eq:test_dist_1_3D}, as seen simply by taking the imaginary part of \eqref{eq:S1_3D_Bashir}. 
	
	Through this analysis of the LKFT for the fermion propagator in 3D, we have established that solutions of $\mathcal{K}_j(s,s';\xi)$ directly from their differential equations satisfy group properties postulated in Section \ref{ss:LKFT_group} and agree with the general solution obtained in Section \ref{ss:LKFT_spectral_rep}. In Bashir and Raya \cite{Bashir:2002sp}, LKFT for the fermion propagator is solved through Fourier transforms to and from coordinate space assuming the propagator is the free-particle one in the Landau gauge. Therefore (16, 17) of Bashir and Raya\cite{Bashir:2002sp} only apply under this assumption. However, through the spectral representation, any propagator function can be represented as a linear combination of free-particle propagators with different mass. Since the LKFT is also linear, results in Bashir and Raya\cite{Bashir:2002sp} can be generalized to accommodate any initial conditions having spectral representations themselves. Consequently, \eqref{eq:K_1_free_prop_4D} holds regardless of the assumed behavior of the propagator in the initial gauge. In our reduction of the exact solution to LKFT in any dimensions by \eqref{eq:k_exponential} and \eqref{eq:def_phi} to the special case of 3D, gauge covariance of the fermion propagator is solved directly in Minkowski momentum space through the language of spectral representation, and therefore is independent of the initial conditions specified in any one gauge.
	\section{LKFT for fermion propagator in 4D\label{ss:LKFT_fermion_4D}}
	The group properties of LKFT for the fermion propagator spectral functions are maintained by \eqref{eq:k_exponential}, for any positive $\epsilon$. However when $d\rightarrow 4$, only the leading expansions in $\epsilon$ are required. Therefore one expects distributions $\mathcal{K}_j$ to become simpler than \eqref{eq:kn_free_prop} in this particular limit, as they did in 3D. However, to obtain the correct expansions, knowledge of the divergent part for the fermion propagator is required. By analyzing divergences alone, the LKFT specifies that the fermion propagator wavefunction renormalization \cite{Johnson:1959zz,Sonoda:2000kn} is $Z_2(\xi)=Z_2(0)\exp\left[-\alpha\xi/(4\pi\epsilon)\right]$. Additionally, there is no ultraviolet divergence for the fermion self-energy in the Landau gauge, so we can take $Z_2(0)=1$. After dimensional regularization, any term at $\mathcal{O}(\epsilon^1)$ is regarded as higher order in the exponential. When \eqref{eq:kn_free_prop} is convergent, the LKFT for the fermion propagator in 4D is found once the proper limit of $\epsilon\rightarrow 0$ is taken.
	
	To proceed to evaluating \eqref{eq:kn_free_prop} with small $\epsilon$, consider the original definition of the Gamma function 
	\begin{equation}
	\Gamma(s)=\int_{0}^{+\infty}dx~x^{s-1}e^{-x}.\label{eq:def_Gamma}
	\end{equation}
	After reparameterizing Gamma functions in the numerator of the double series expansion of \eqref{eq:kn_zbeta}, we obtain
	\begin{align}
	& \quad \sum_{m=0}^{+\infty}\dfrac{\Gamma(n+\beta+(m-1)\epsilon-1)\Gamma(\beta+m\epsilon)}{\Gamma(n+\beta-\epsilon-1)\Gamma(\beta)}\dfrac{(-\overline{\alpha}z^\epsilon)^m}{m!}\nonumber\\
	& =\sum_{m=0}^{+\infty}\int_{0}^{+\infty}dx\int_{0}^{+\infty}dy\dfrac{e^{-x-y}}{\Gamma(n+\beta-\epsilon-1)\Gamma(\beta)}x^{n+\beta-\epsilon-2}y^{\beta-1}\dfrac{[-(xyz)^\epsilon\overline{\alpha}]^m}{m!}\nonumber\\
	& =\int_{0}^{+\infty}dx\int_{0}^{+\infty}dy\dfrac{x^{n+\beta-\epsilon-2}y^{\beta-1}}{\Gamma(n+\beta-\epsilon-1)\Gamma(\beta)}\exp\left[-x-y-\overline{\alpha}\left(xyz\right)^\epsilon\right] \label{eq:kn_zbeta_para}
	\end{align}
	Equation~\eqref{eq:kn_zbeta_para} is an alternative to \eqref{eq:kn_zbeta}. 
	For any fermion propagator function $S_j(p^2)$ in 4D, having established that its divergent part is merely $Z_2=e^{-\alpha\xi/(4\pi\epsilon)}$, 
	it must also be the only divergence for $\mathcal{K}_nz^\beta$ for small $\epsilon$. With the renormalization factor in mind, $\mathcal{K}_nz^\beta$ is properly renormalized once its logarithm is truncated to $\mathcal{O}(\epsilon^0)$. Furthermore, integrals over parameters $x$ and $y$ do not modify the $1/\epsilon$ divergences of $\mathcal{K}_nz^\beta$ because the integral definition of the Gamma function by \eqref{eq:def_Gamma} extends into the complex plane. Then for each integral element of \eqref{eq:kn_zbeta_para},
	\begin{align}
	& \quad \ln\Bigg\{ \dfrac{x^{n+\beta-\epsilon-2}y^{\beta-1}}{\Gamma(n+\beta-\epsilon-1)\Gamma(\beta)}\exp[-x-y-\overline{\alpha}(xyz)^\epsilon]\Bigg\}\nonumber\\
	& =-x-y+(n+\beta-\epsilon-2)\ln x+(\beta-1)\ln y\nonumber\\
	& \quad -\dfrac{\alpha\xi}{4\pi}\dfrac{\Gamma(\epsilon)\Gamma(1-\epsilon)}{\Gamma(1+\epsilon)}\left(\dfrac{4\pi\mu^2}{p^2}xyz \right)^\epsilon -\ln\Gamma(n+\beta-\epsilon-1)-\ln\Gamma(\beta)\nonumber\\
	& =-x-y+(n+\beta-2)\ln x+(\beta-1)\ln y-\ln\Gamma(n+\beta-1)\nonumber\\
	& \quad -\ln\Gamma(\beta)-\dfrac{\alpha\xi}{4\pi}\left[\dfrac{1}{\epsilon}+\gamma_E+\ln\left(\dfrac{4\pi\mu^2}{p^2}xyz \right) \right]+\mathcal{O}(\epsilon^1).
	\end{align}
	After regularization,
	\begin{align}
	z^{-\beta}\mathcal{K}_j z^\beta & =\dfrac{(\mu^2z/p^2)^{-\alpha\xi/(4\pi)}}{\Gamma(n+\beta-1)\Gamma(\beta)}\exp\left[-\dfrac{\alpha\xi}{4\pi}\left(\dfrac{1}{\epsilon}+\gamma_E+\ln 4\pi+\mathcal{O}(\epsilon^1)\right) \right]\times\nonumber\\
	&\quad  \int_{0}^{+\infty}dx\int_{0}^{+\infty}dye^{-x-y}x^{n+\beta-2-\alpha\xi/(4\pi)}y^{\beta-1-\alpha\xi/(4\pi)}\nonumber\\
	& =\dfrac{\Gamma\left(n+\beta-1-\dfrac{\alpha\xi}{4\pi}\right)\Gamma\left(\beta-\dfrac{\alpha\xi}{4\pi}\right)}{\Gamma(n+\beta-1)\Gamma(\beta)}\left(\dfrac{\mu^2z}{p^2} \right)^{-\alpha\xi/(4\pi)}\times\nonumber\\
	&\quad \exp\left[-\dfrac{\alpha\xi}{4\pi}\left(\dfrac{1}{\epsilon}+\gamma_E+\ln 4\pi+\mathcal{O}(\epsilon^1) \right) \right].
	\end{align}
	While from the series definition of hypergeometric functions,
	\begin{align}
	& \quad \sum_{\beta=1}^{+\infty}\dfrac{\Gamma\left(n+\beta-1-\dfrac{\alpha\xi}{4\pi}\right)\Gamma\left(\beta-\dfrac{\alpha\xi}{4\pi}\right)}{\Gamma(n+\beta-1)\Gamma(\beta)}z^\beta\nonumber\\
	& =\dfrac{z}{\Gamma(n)}\Gamma\left(n -\dfrac{\alpha\xi}{4\pi}\right) \Gamma\left( 1-\dfrac{\alpha\xi}{4\pi}\right)\sum_{\beta=0}^{+\infty}\dfrac{\Gamma\left(n+\beta-\dfrac{\alpha\xi}{4\pi}\right)\Gamma\left(\beta+1 -\dfrac{\alpha\xi}{4\pi}\right)}{\Gamma\left(n -\dfrac{\alpha\xi}{4\pi}\right)\Gamma\left(1-\dfrac{\alpha\xi}{4\pi}\right)}\dfrac{\Gamma(n)}{\Gamma(n+\beta)}\dfrac{z^\beta}{\beta!}\nonumber\\
	& =\dfrac{z}{\Gamma(n)}\Gamma\left(n -\dfrac{\alpha\xi}{4\pi}\right) \Gamma\left( 1-\dfrac{\alpha\xi}{4\pi}\right)~_2F_1\left(1-\dfrac{\alpha\xi}{4\pi},n-\dfrac{\alpha\xi}{4\pi};n;z\right)
	\end{align}
	Therefore for $\mathcal{K}_j$ acting on the free-particle propagator, \eqref{eq:kn_free_prop} becomes
	\begin{align}
	\mathcal{K}_j\dfrac{1}{p^2-s+i\varepsilon} & =\dfrac{-1}{p^2}\left(\dfrac{\mu^2z}{p^2} \right)^{-\alpha\xi/(4\pi)}\exp\left[-\dfrac{\alpha\xi}{4\pi}\left(\dfrac{1}{\epsilon}+\gamma_E+\ln 4\pi+\mathcal{O}(\epsilon^1) \right)\right]\times\nonumber\\
	&\quad  \sum_{\beta=1}^{+\infty}\dfrac{\Gamma\left(n+\beta-\dfrac{\alpha\xi}{4\pi}\right)\Gamma\left(\beta-\dfrac{\alpha\xi}{4\pi}\right)}{\Gamma(n+\beta-1)\Gamma(\beta)}z^\beta\nonumber\\
	& =\dfrac{-z}{p^2\Gamma(n)}\left(\dfrac{\mu^2z}{p^2} \right)^{-\alpha\xi/(4\pi)}\exp\left[-\dfrac{\alpha\xi}{4\pi}\left(\dfrac{1}{\epsilon}+\gamma_E+\ln 4\pi+\mathcal{O}(\epsilon^1) \right)\right]\times\nonumber\\
	& \quad \Gamma\left(n -\dfrac{\alpha\xi}{4\pi}\right) \Gamma\left( 1-\dfrac{\alpha\xi}{4\pi}\right)~_2F_1\left(1-\dfrac{\alpha\xi}{4\pi},n-\dfrac{\alpha\xi}{4\pi};n;z\right).\label{eq:LKFT_rho_SF_4D}
	\end{align}
	This is our general result. In the special case when the propagator is assumed to be free in the Landau gauge, \textit{i.e.} $\rho_j$ are $\delta$-functions, \eqref{eq:LKFT_rho_SF_4D} reduces to the results found by Bashir and Raya\cite{Bashir:2002sp} up to differences in renormalization schemes.
	
	To generate \eqref{eq:LKFT_rho_SF_4D} from the free-particle propagator, consider a positive change in $\xi$. Naturally adopting the convention  $\nu=\alpha\xi/(4\pi)$ used in Bashir and Raya\cite{Bashir:2002sp}, for instance, we have from \eqref{eq:Dc2Fa}
	\begin{align}
	& I^{\nu}z^{-\nu}~_2F_1(1,n;n;z)=\Gamma(1-\nu)~_2F_1(1-\nu,n;n;z),\\
	& I^{\nu}z^{n-1-\nu}~_2F_1(1-\nu,n;b;z)=\dfrac{\Gamma(n-\nu)}{\Gamma(n)}z^{n-1}~_2F_1(1-\nu,n-\nu;n;z),
	\end{align}
	since the hypergeometric $_2F_1(a,b;c;z)$ is symmetric in parameters $a$ and $b$, and	where the fractional differential operation in \eqref{eq:Dc2Fa} becomes fractional integration for positive $\xi$. Combining these two identities gives
	\begin{equation}
	z^{n-1}I^{\nu}z^{n-1-\nu}I^{\nu}z^{-\nu}~_2F_1(1,n;n;z)=\dfrac{\Gamma(n-\nu)\Gamma(1-\nu)}{\Gamma(n)}~_2F_1(1-\nu,n-\nu;n;z).\label{eq:LKFT_SF_4D_generation}
	\end{equation}
	After representing the free-particle propagator as a hypergeometric function using \eqref{eq:free_particle_prop_2F1}, comparing \eqref{eq:LKFT_rho_SF_4D} with \eqref{eq:LKFT_SF_4D_generation} yields
	\begin{equation}
	\mathcal{K}_j(\xi)=\left(\dfrac{\mu^2z}{p^2} \right)^{-\nu}\exp\bigg\{-\nu\left[\dfrac{1}{\epsilon}+\gamma_E+\ln 4\pi+\mathcal{O}(\epsilon^1) \right] \bigg\}z^{2-n}I^{\nu}z^{n-1-\nu}I^{\nu}z^{-\nu-1},\label{eq:kn_small_epsilon}
	\end{equation}
	again with $n=3,~2$ for $j=1,~2$. While for negative $\xi$, the fractional integration operators $I^\nu$ are replaced by derivative operators $D^{-\nu}=D^{|\nu|}$. \eqref{eq:kn_small_epsilon} then becomes 
	\begin{equation}
	\mathcal{K}_j(-\xi)=\left(\dfrac{\mu^2z}{p^2} \right)^{-\nu}\exp\bigg\{-\nu\left[\dfrac{1}{\epsilon}+\gamma_E+\ln 4\pi+\mathcal{O}(\epsilon^1) \right] \bigg\}z^{2-n}D^{-\nu}z^{n-1-\nu}D^{-\nu}z^{-\nu-1}.\label{eq:kn_small_epsilon_inverse}
	\end{equation}	
	One can verify that from (\ref{eq:kn_small_epsilon},~\ref{eq:kn_small_epsilon_inverse}), by acting on $z^\beta$, that $\mathcal{K}_j(\xi_1)\mathcal{K}_j(\xi_2)=\mathcal{K}_j(\xi_1+\xi_2)$ and $\mathcal{K}_j^{-1}(\xi)=\mathcal{K}_j(-\xi)$. Therefore the simplified form of LKFT for fermion propagator spectral functions in 4D also maintains group properties explicitly.
	\section{Summary and conclusions\label{ss:conclusions}}
Working in covariant gauges we have shown here that the Landau-Khalatnikov-Fradkin transformation (LKFT) defines a group of transformations parametrized by the gauge label $\xi$. These transformations define how a  propagator in one covariant gauge is related to that in any other. These transformations are readily studied if we assume the propagator satisfies a spectral reprsentation.  As an explicit example  we have investigated the fermion propagator in QED, which is expected to have the analytic properties required for such a representation. The LKFT then demands the spectral functions  obey exact  transformation properties to be gauge covariant. These hold in any dimension $d < 4$, naturally involving fractional calculus. In three dimensions when the calculus is of integer order, 
we show how our results generalize those obtained earlier in a special case by Bashir and Raya \cite{Bashir:2002sp}. As we approach four dimensions, the general results can be expanded in powers of $\epsilon = 2-d/2$.
The complexity of fractional calculus then becomes apparent. The solutions inevitably involve distributions with fractional orders of delta and theta-functions. Nevertheless, considering arbitrary (non-integer) dimensions provides insights into how gauge covariance connects the properties of field theory Green's functions in different dimensions.

What constraints gauge covariance imposes on truncations of the Schwinger-Dyson equations for these same Green's functions (propagators and vertices) is the subject of on-going study.
\begin{acknowledgments}
This material is based upon work supported by the U.S. Department of Energy, Office of Science, Office of Nuclear Physics under contract DE-AC05-06OR23177 that funds Jefferson Lab research. The authors would like to thank Professor Keith Ellis and other members of the Institute for Particle Physics Phenomenology (IPPP) of Durham University for kind hospitality during their visit when this article was finalized.
\end{acknowledgments}

\appendix
	\section{Evaluating Loop Integrals in Minkowski Space \label{ss:loop_Minkowski}}
	For a given loop integral in quantum field theory, after Feynman parameterization, one possible form of the integral is,
	\begin{equation}
	L_{0n}(\Delta,\epsilon)=\int d\underline{l}\dfrac{1}{(l^2-\Delta+i\varepsilon)^n},
	\end{equation}
	with $d\underline{l}\equiv d^d l/(2\pi)^d$.
	where $\Delta$ is the mass function for the combined denominator, and $\varepsilon$ denotes the Feynman prescription for timelike integrals.
	The textbook version of evaluating $L_{0n}$ is to apply Wick rotation directly as $l_0=il_4$, then evaluate $L_{0n}$ using dimensional regularization (or other regularization schemes). We want to explore the possibility of evaluating loop integrals directly in Minkowski space without Wick rotation, while still employing dimensional regularization. 

	Since $l^2=l_0^2-\overrightarrow{l}^2=l_0^2-\overrightarrow{l}\cdot \overrightarrow{l}$, where $l_0$ is the temporal component of loop momentum while $\overrightarrow{l}$ represents all spatial components. The number of components described by $\overrightarrow{l}$ is related to the number of spacetime dimensions. We take the convention that dimensional regularization is only allowed to change spatial dimensions, leaving the temporal component alone. 
	When evaluating the contour for the temporal component of loop integral, the Feynman prescription tells us that when the contour is closed above, only the $l_0=-E_{\overrightarrow{l}}$ pole is included, the residue of which is the result of the temporal integral. As expected, an identical result is obtained if instead the contour is closed from below, encircling the pole at $l_0=+E_{\overrightarrow{l}}$. 
	With Wick rotation $l_0=il_4$, one can easily verify that the contour for temporal integration is rotated $90^{\circ}$ counterclockwise around the origin, rendering the same pole encompassed in the contour as required by Feynman prescription for Minkowski space temporal integrals, therefore producing identical results.
	
	To see how to evaluate $L_{0n}$ in Minkowski space directly, consider its temporal integration first. Because contributions from the infinite radius arc vanish for large enough $n$,
	\begin{equation}
	\int dl_0\dfrac{1}{(l_0^2-E_{\overrightarrow{l}}^2)^n}=2\pi i~\mathrm{Res}_{l_0\rightarrow -E_{\overrightarrow{l}}}\dfrac{1}{(l_0+E_{\overrightarrow{l}})^n(l_0-E_{\overrightarrow{l}})^n},
	\end{equation}
	where $E_{\overrightarrow{l}}=\sqrt{\overrightarrow{l}^2+\Delta}$. Next, since the order of the pole at $-E_{\overrightarrow{l}}$ is $n$,
	\begin{align}
	& \quad \mathrm{Res}_{l_0\rightarrow -E_{\overrightarrow{l}}}\dfrac{1}{(l_0+E_{\overrightarrow{l}})^n(l_0-E_{\overrightarrow{l}})^n}\nonumber\\
	& =\lim\limits_{l_0\rightarrow -E_{\overrightarrow{l}}}\dfrac{1}{(n-1)!}\left(\dfrac{d}{dl_0} \right)^{n-1}(l_0-E_{\overrightarrow{l}})^{-n}\nonumber\\
	& =(-1)^{n}2^{-2n+1}\dfrac{\Gamma(2n-1)}{[\Gamma(n)]^2E_{\overrightarrow{l}}^{2n-1}}.
	\end{align}
	While for the spatial integration, dimensional regularization is applied such that 
	\[\int d\overrightarrow{l}=\int d\Omega_{d-1}\int_{0}^{+\infty}d|\overrightarrow{l}|~|\overrightarrow{l}|^{d-2},\]
	where $|\overrightarrow{l}|=\sqrt{\overrightarrow{l}^2}$ and for spherical symmetric kernels $\int d\Omega_{d-1}=2\pi^{(d-1)/2}/\Gamma((d-1)/2)$.
	Therefore
	\begin{align}
	L_{0n}(\Delta,\epsilon)& =\dfrac{1}{(2\pi)^d}\int d\overrightarrow{l}\int dl_0\dfrac{1}{(l_0^2-E_{\overrightarrow{l}}^2)^n}\nonumber\\
	& =\dfrac{2\pi i}{(2\pi)^d}\int d\overrightarrow{l}\dfrac{(-1)^n 2^{-2n+1}\Gamma(2n-1)}{[\Gamma(n)]^2\left(\overrightarrow{l}^2+\Delta\right)^{n-1/2}}\nonumber\\
	& =\dfrac{i(-1)^n2^{-2n+1}\Gamma(2n-1)}{(2\pi)^{d-1}[\Gamma(n)]^2}\dfrac{\pi^{(d-1)/2}}{\Gamma\left(\dfrac{d-1}{2} \right)}\int_{0}^{+\infty}d\overrightarrow{l}^2\dfrac{\left(\overrightarrow{l}^2\right)^{(d-3)/2}}{\left(\overrightarrow{l}^2+\Delta\right)^{n-1/2}}.
	\end{align} 
	Substituting $x=\left(\overrightarrow{l}^2/\Delta+1 \right)^{-1}$ for the integration variable, we have
	\begin{equation}
	L_{0n}(\Delta,\epsilon)=\dfrac{i(-1)^n2^{-2n+2-d}\Gamma(2n-1)}{\pi^{(d-1)/2}[\Gamma(n)]^2\Delta^{n-d/2}}\dfrac{1}{\Gamma\left(\dfrac{d-1}{2} \right)}\int_{0}^{1}dx~x^{n-d/2-1}(1-x)^{(d-1)/2-1}.
	\end{equation}
	This integral over $x$ is then just the Euler Beta function $B(n-d/2,(d-1)/2)$.
	Noting that
	\[\dfrac{\Gamma(2n-1)}{\Gamma(n)\Gamma(n-1/2)}=\dfrac{2^{2n-2}}{\sqrt{\pi}}, \]
	we arrive at
	\begin{equation}
	L_{0n}(\Delta,\epsilon)=\dfrac{i(-1)^n}{(4\pi)^{d/2}}\dfrac{\Gamma(n-d/2)}{\Gamma(n)\Delta^{n-d/2}},
	\end{equation}
	which agrees with (A.44) of Peskin and Schroeder \cite{peskin1995introduction} for the Wick-rotated result.
	
	The more general integrals ${L_{mn}(\Delta,\epsilon)=\int d\underline{l}~l^{2m}/(l^2-\Delta+i\varepsilon)^n}$ are determined by combinations of $L_{0r}(\Delta,\epsilon)$. Consequently, the result for $L_{mn}(\Delta,\epsilon)$ is 
	\begin{equation}
	L_{mn}(\Delta,\epsilon)=\dfrac{i(-1)^{n-m}}{(4\pi)^{d/2}}\dfrac{\Gamma(n-d/2-m)}{\Gamma(n)\Delta^{n-d/2-m}}\prod_{m'=1}^{m}\left(\dfrac{d}{2}+m'-1\right),
	\end{equation}
	with $m\geq 1$ and $n\geq m+1$ to ensure the convergence of the $l_0$ integral. This also agrees with Peskin and Schroeder \cite{peskin1995introduction}. 

	While in the special case of $\Delta=0$, singularities of $l_0$ integrals are modified from the case of $\Delta\neq 0$. Therefore integrations for the massless case require a separate discussion, which is not needed in this article.
	\section{$\Xi_j(p^2,s)$ with $d=4-2\epsilon$ as hypergeometric functions\label{ss:Xi_12_epsilon}}
	From the Euler type integral definition of hypergeometric functions \cite{abramowitz1964handbook}
	\begin{equation}
	\int_{0}^{1}dx~x^{b-1}(1-x)^{c-b-1}(1-zx)^{-a}=\dfrac{\Gamma(b)\Gamma(c-b)}{\Gamma(c)}~_2F_1(a,b;c;z),
	\end{equation}
	we express the following two integrals as hypergeometric functions,
	\begin{align}
	I_0(z,\epsilon)& \equiv \int_{0}^{1}dx~\dfrac{2x}{(1-x)^\epsilon (1-xz)^\epsilon}=\dfrac{2~_2F_1(\epsilon,2;3-\epsilon;z)}{(1-\epsilon)(2-\epsilon)},\\
	I_1(z,\epsilon)& \equiv \int_{0}^{1}dx~\dfrac{2x}{(1-x)^\epsilon (1-xz)^{1+\epsilon}}=\dfrac{2~_2F_1(\epsilon+1,2;3-\epsilon;z)}{(2-\epsilon)(1-\epsilon)}\nonumber\\
	& =\dfrac{-2}{(1-\epsilon)(z-1)}+\dfrac{2[1-\epsilon(z+1)]}{(2-\epsilon)(1-\epsilon)(z-1)}~_2F_1(1,1+\epsilon;3-\epsilon;z).
	\end{align}
	Applying this result to (\ref{eq:def_Xi_1},~\ref{eq:def_Xi_2}) gives
	\begin{align}
	\Xi_1(p^2,s)& =\Gamma(\epsilon)\left(\dfrac{4\pi\mu^2}{s} \right)^\epsilon [(1-\epsilon)I_0(z,\epsilon)+\epsilon I_1(z,\epsilon)]\nonumber\\
	\Xi_2(p^2,s)& =\Gamma(\epsilon)\left(\dfrac{4\pi\mu^2}{s} \right)^\epsilon\left[(1-\epsilon)I_0(z,\epsilon)+\dfrac{\epsilon(z+1)}{2}I_1(z,\epsilon)\right].
	\end{align}
	Using (15.2.10) in Abramowitz and Stegun\cite{abramowitz1964handbook}, also listed in Appendix \ref{ss:identities_2F1}, with $a=\epsilon+1,~b=2,~c=3-\epsilon$, we obtain,
	\begin{align*}
	& \quad (1-\epsilon)~_2F_1(\epsilon,2;3-\epsilon;z)\nonumber\\
	&=-\dfrac{\epsilon+1}{2}(z-1)~_2F_1(\epsilon+2,2;3-\epsilon;z)-\dfrac{1}{2}[3\epsilon-1+(1-\epsilon)z]~_2F_1(1+\epsilon,2;3-\epsilon;z).
	\end{align*}
	While applying (15.2.14) and (15.2.17) with $a=\epsilon+1,~b=2$ and $c=3-\epsilon$ respectively, we have, 
	\begin{align*}
	&(1-\epsilon)~_2F_1(\epsilon+1,2;3-\epsilon;z)+(\epsilon+1)~_2F_1(\epsilon+2,2;3-\epsilon;z) =2~_2F_1(\epsilon+1,3;3-\epsilon;z)\\
	&(1-2\epsilon)~_2F_1(\epsilon+1,2;3-\epsilon;z)+(\epsilon+1)~_2F_1(\epsilon+2,2;3-\epsilon;z)=(2-\epsilon)~_2F_1(\epsilon+1,2;2-\epsilon;z).
	\end{align*}
	Therefore
	\begin{align*}
	&\quad (1-\epsilon)~_2F_1(\epsilon,2;3-\epsilon;z)+\epsilon~_2F_1(\epsilon+1,2;\epsilon-2;z)\nonumber\\
	& =-\dfrac{\epsilon+1}{2}(z-1)~_2F_1(\epsilon+2,2;3-\epsilon;z)-\dfrac{(1-\epsilon)}{2}(z-1)~_2F_1(\epsilon+1,2,3-\epsilon,z),\\
	& =(1-z)~_2F_1(\epsilon+1,3;3-\epsilon;z),
	\end{align*}
	and
	\begin{align*}
	& \quad (1-\epsilon)~_2F_1(\epsilon,2;3-\epsilon;z)+\dfrac{\epsilon}{2}(z+1)~_2F_1(\epsilon+1,2;3-\epsilon;z)\nonumber\\
	& =-\dfrac{\epsilon+1}{2}(z-1)~_2F_1(\epsilon+2,2;3-\epsilon;z)-\dfrac{1-2\epsilon}{2}(z-1)~_2F_1(\epsilon+1,2;3-\epsilon;z)\\
	& =(1-z)\dfrac{2-\epsilon}{2}~_2F_1(\epsilon+1,2;2-\epsilon;z).
	\end{align*}
	Then the $z$ dependences of $\Xi_j/(p^2-s)$ combine as
	\begin{align}
	\dfrac{\Xi_1}{p^2-s}& =\dfrac{\Gamma(\epsilon)}{s(z-1)}\left(\dfrac{4\pi\mu^2}{s}\right)^\epsilon\dfrac{2}{(1-\epsilon)(2-\epsilon)}\bigg\{(1-\epsilon)~_2F_1(\epsilon,2;3-\epsilon;z)\nonumber\\
	& \quad\hspace{4.5cm} +\epsilon ~_2F_1(\epsilon+1,2;3-\epsilon;z)\bigg\}\nonumber\\
	& =\dfrac{\Gamma(\epsilon)}{s}\left(\dfrac{4\pi\mu^2}{s}\right)^\epsilon\dfrac{-2}{(1-\epsilon)(2-\epsilon)}~_2F_1(\epsilon+1,3;3-\epsilon;z)\\
	\dfrac{\Xi_2}{p^2-s}&=\dfrac{\Gamma(\epsilon)}{s(z-1)}\left(\dfrac{4\pi\mu^2}{s}\right)^\epsilon\dfrac{2}{(1-\epsilon)(2-\epsilon)}\bigg\{(1-\epsilon)~_2F_1(\epsilon,2;3-\epsilon;z)\nonumber\\
	& \quad\hspace{4.5cm} +\dfrac{\epsilon(z+1)}{2} ~_2F_1(\epsilon+1,2;3-\epsilon;z)\bigg\}\nonumber\\
	& =\dfrac{\Gamma(\epsilon)}{s}\left(\dfrac{4\pi\mu^2}{s}\right)^\epsilon\dfrac{-1}{1-\epsilon}~_2F_1(\epsilon+1,2;2-\epsilon;z).
	\end{align}
	Using results in Appendix \ref{ss:identities_2F1}, one can verify that \eqref{eq:Xi_1_reduced} and \eqref{eq:Xi_2_reduced} reduce to results by direct calculation of integrations over Feynman parameters after taking the $\epsilon=1/2$ limit and the $\epsilon\rightarrow 0$ expansion, respectively. 
	\section{Useful identities for hypergeometric functions $~_2F_1(a,b;c;z)$\label{ss:identities_2F1}}
	We collect identities we have used from Abramowitz and Stegun \cite{abramowitz1964handbook}.
	\subsection{Notations and definitions}
	\begin{equation}
	~_2F_1(a,b;c;z)=~_2F_1(a,b,c,z)=F(a,b;c;z)=\sum_{n=0}^{+\infty}\dfrac{(a)_n (b)_b}{(c)_n n!}z^n,\label{eq:2F1_Taylor}
	\end{equation}
	where $(a)_n$ is the Pochharmer symbol given by
	\begin{equation}
	(a)_n=a(a+1)(a+2)\dots (a+n-1)=\dfrac{\Gamma(a+n)}{\Gamma(a)}.
	\end{equation} 
	Additionally, the Gamma function definition of Pochharmer symbol applies even when $n$ is not an integer.
	\subsection{Identities for $_2F_1(a,b;c;z)$\label{ss:identities_2F1_AS}}
	Identities listed in this subsection are selected equations from Abramowitz and Stegun \cite{abramowitz1964handbook}. Equations numbered from the left are labeled by their the original numbers.
	\paragraph{Special Elementary Cases of Gauss Series}
	\begin{align*}
	& (15.1.4)\quad F\left(\dfrac{1}{2},1;\dfrac{3}{2};z^2\right)=\dfrac{1}{2z}\ln\left(\dfrac{1+z}{1-z}\right)=\dfrac{\mathrm{arctanh}(z)}{z}\\
	& (15.1.5)\quad F\left(\dfrac{1}{2},1;\dfrac{3}{2};z^2\right)=\dfrac{\mathrm{arctan}(z)}{z}\\
	& (15.1.8)\quad F(a,b;b;z)=(1-z)^{-a}\\
	\end{align*}
	\paragraph{Differentiation Formulas}
	\begin{align*}
	& (15.2.3)\quad \dfrac{d^n}{dz^n}[z^{a+n-1}F(a,b;c;z)]=(a)_n z^{a-1}F(a+n,b;c;z)\\
	& (15.2.4)\quad \dfrac{d^n}{dz^n}[z^{c-1}F(a,b;c;z)]=(c-n)_n z^{c-n-1}F(a,b;c-n;z)\\
	\end{align*}
	\paragraph{Gauss' relations for contiguous functions}
	\begin{align*}
	& (15.2.10)\quad (c-a)F(a-1,b;c;z)+(2a-c-az+bz)F(a,b;c;z)+a(z-1)F(a+1,b;c;z)=0\\
	& (15.2.14)\quad (b-a)F(a,b;c;z)+aF(a+1,b;c;z)-bF(a,b+1;c;z)=0\\
	& (15.2.17)\quad (c-a-1)F(a,b;c;z)+aF(a+1,b;c;z)-(c-1)F(a,b;c-1;z)=0
	\end{align*}
	\paragraph{Integral Representations and Transformation Formulas}
	\begin{equation*}
	(15.3.1)\quad F(a,b;c;z)=\dfrac{\Gamma(c)}{\Gamma(b)\Gamma(c-b)}\int_{0}^{1}dt~t^{b-1}(1-t)^{c-b-1}(1-tz)^{-a}
	\end{equation*}
	with $\mathrm{Re}\{c\}>\mathrm{Re}\{b\}>0$.
	\begin{align}
	(15.3.5)\quad F(a,b;c;z)& =(1-z)^{-b}F(b,c-a;c;z/(z-1))\nonumber\\
	(15.3.6)\quad F(a,b;c;z)& =\dfrac{\Gamma(c)\Gamma(c-a-b)}{\Gamma(c-a)\Gamma(c-b)}F(a,b;a+b-c+1;1-z)\nonumber\\
	& \quad +(1-z)^{c-a-b}\dfrac{\Gamma(c)\Gamma(a+b+c)}{\Gamma(a)\Gamma(b)}F(c-a,c-b;c-a-b+1;1-z),\label{eq:2F1_singular_z1}
	\end{align}
	with $c-a-b\notin \mathbf{N}$. When $b-c=m\in \mathbf{N}^*$,
	\begin{align*}
	(15.3.14)&\quad F(a,a+m;c;z) =F(a+m,a;c;z)\\
	& =\dfrac{\Gamma(c)(-z)^{-a-m}}{\Gamma(a+m)\Gamma(c-a)}\sum_{n=0}^{+\infty}\dfrac{(a)_{n+m}(1-c+a)_{n+m}}{n!(n+m)!}z^{-n}\big\{\ln(-z)\\
	& \quad +\psi(1+m+n)+\psi(1+n)-\psi(a+m+n)-\psi(c-a-m-n)\big\}\\
	& \quad+ (-z)^{-a}\dfrac{\Gamma(c)}{\Gamma(a+m)}\sum_{n=0}^{m-1}\dfrac{\Gamma(m-n)(a)_n}{n!\Gamma(c-a-n)}z^{-n}\\
	& \quad (\mathrm{for}~|\mathrm{arg}(-z)|<\pi,~|z|>1,~(c-a)\neq \mathbf{Z}).
	\end{align*}
	\subsection{Leading expansions on small parameters\label{ss:identities_2F1_epsilon}}
	The definition of derivative on parameters
	\begin{equation}
	~_2F_1^{(l,m,n,0)}(\alpha,\beta;\gamma;z)\equiv\lim\limits_{(a,b,c)\rightarrow (\alpha,\beta,\gamma)}\dfrac{\partial^{l+m+n}}{\partial a^l\partial b^m\partial c^n}~_2F_1(a,b;c;z).
	\end{equation}
	For the purpose of calculating $\epsilon\rightarrow 0$ limits, only first order derivatives are required on parameters. One simple example that is relevant to the $\epsilon\rightarrow 0$ limit of the LKFT is
	\begin{equation}
	~_2F_1^{(1,0,0,0)}(1,n;n;z)=\lim\limits_{a\rightarrow 1}\dfrac{\partial}{\partial a}(1-z)^{-a}=-\dfrac{\ln(1-z)}{1-z}.
	\end{equation}
	
	A straightforward way to calculate these leading derivatives is to use the series definition given by \eqref{eq:2F1_Taylor}. First, consider the derivative of the Pochhammer symbol
	\begin{equation}
	\dfrac{\partial}{\partial a}(a)_n=\dfrac{\partial}{\partial a}\dfrac{\Gamma(a+n)}{\Gamma(a)}=\dfrac{\Gamma(a+n)}{\Gamma(a)}\left[\dfrac{\partial}{\partial a}\ln\Gamma(a+n)-\dfrac{\partial}{\partial a}\ln\Gamma(a) \right]=(a)_n\left[\psi(a+n)-\psi(a)\right],
	\end{equation}
	where $\psi(z)=d\ln\Gamma(z)/dz$ is the digamma function, and
	\begin{equation}
	\psi(z+1)=\psi(z)+1/z.
	\end{equation}
	For integer $n$, $\psi(n)=H_{n-1}-\gamma_E$, where the harmonic number is defined by ${H_{n-1}=\sum_{m=1}^{n-1}\frac{1}{m}}$. Then
	\begin{equation}
	\psi(a+n)-\psi(a)=\dfrac{1}{a+n-1}+\dfrac{1}{a+n-2}+\dots +\dfrac{1}{a}=\sum_{m=0}^{n-1}\dfrac{1}{a+m},\quad \mathrm{for}~n\in \mathbf{N^*}.
	\end{equation}
	
	To calculate $~_2F_1^{(0,0,1,0)}(1,3;3;z)$ and $~_2F_1^{(0,0,1,0)}(1,2;2;z)$, consider the following series expansion:
	\begin{equation}
	\lim\limits_{c\rightarrow b}\dfrac{\partial}{\partial c}~_2F_1(1,b;c;z)=\lim\limits_{c\rightarrow b}\dfrac{\partial}{\partial c}\sum_{n=0}^{+\infty}\dfrac{(b)_n}{(c)_n}z^n=\sum_{n=1}^{+\infty}[\psi(b)-\psi(b+n)]z^n.
	\end{equation}
	Then we have
	\begin{align}
	& ~_2F_1^{(0,0,1,0)}(1,3;3;z)=-\dfrac{z+z^2/2+\ln(1-z)}{z^2(z-1)}\\
	& ~_2F_1^{(0,0,1,0)}(1,2;2;z)=-\dfrac{z+\ln(1-z)}{z(z-1)},
	\end{align}
	from which we finally obtain
	\begin{align}
	& ~_2F_1(1-\epsilon,3;3-\epsilon;z)=\dfrac{-1}{z-1}+\epsilon\left[\dfrac{\ln(1-z)}{z-1}+\dfrac{z+z^2/2+\ln(1-z)}{z^2(z-1)} \right]+\mathcal{O}(\epsilon^1)\\
	& ~_2F_1(1-\epsilon,2;2-\epsilon;z)=\dfrac{-1}{z-1}+\epsilon\left[\dfrac{\ln(1-z)}{z-1}+\dfrac{z+\ln(1-z)}{z(z-1)} \right]+\mathcal{O}(\epsilon^1).
	\end{align}
	\section{Example: the exponent-preserving effect of \eqref{eq:linear_trans_LKFT_epsilon} \label{ss:Example_epn_preserving}}
	Operations constructed to generate $p^2$ dependences from the free-particle propagator using exponent-preserving linear transforms are free from operations on momentum variable $p^2$, an essential criterion for the application of spectral representation of propagators to solve the LKFT. If all operations are exponent-preserving on the variable $z=p^2/s$, after integral variable transform $dz=-p^2s^{-2}ds$ there is no residual $p^2$ multiplication factors. This can be verified by the following example corresponding to the linear transform in \eqref{eq:linear_trans_LKFT_epsilon}. 
	Explicitly,
	\begin{align}
	& \quad z^{\epsilon+1-n}D^\epsilon z^{n-1}D^\epsilon z^\epsilon\nonumber\\
	& =\dfrac{z^{\epsilon+1-n}}{\Gamma(1-\epsilon)}\dfrac{d}{dz}\int_{0}^{z}dz'(z-z')^{-\epsilon}\dfrac{(z')^{n-1}}{\Gamma(1-\epsilon)}\dfrac{d}{dz'}\int_{0}^{z'}dz''(z'-z'')^{-\epsilon}(z'')^\epsilon\nonumber\\
	& =\left(\dfrac{p^2}{s}\right)^{\epsilon+1-n}\dfrac{s^2}{\Gamma(1-\epsilon)p^2}\dfrac{d}{ds}\int_{s}^{+\infty}ds'\dfrac{p^2}{(s')^2}\left(\dfrac{p^2}{s}-\dfrac{p^2}{s'} \right)^{-\epsilon}\left(\dfrac{p^2}{s'} \right)^{n-1}\times\nonumber\\
	&\quad{\hspace{4cm}} \dfrac{(s')^2}{\Gamma(1-\epsilon)p^2}\dfrac{d}{ds'}\int_{s'}^{+\infty}ds''\dfrac{p^2}{(s'')^2}\left(\dfrac{p^2}{s'}-\dfrac{p^2}{s''}\right)^{-\epsilon}\left(\dfrac{p^2}{s''}\right)^\epsilon\nonumber\\
	&= \dfrac{s^{1+n-\epsilon}}{(\Gamma(1-\epsilon))^2}\dfrac{d}{ds}\int_{s}^{+\infty}ds'(s')^{1-n+\epsilon}\left(\dfrac{s'}{s}-1 \right)^{-\epsilon}\dfrac{d}{ds'}\int_{s'}^{+\infty}ds''(s'')^{-2}\left(\dfrac{s''}{s'}-1\right)^{-\epsilon},
	\end{align}
	which being exponent-preserving is independent of $p^2$.
	\section{Properties of the distribution $\mathcal{K}_1$ in 3D}
	\subsection{As the solution to its differential equation\label{ss:k1_3D_diff}}
	Apparently \eqref{eq:test_dist_1_3D} reduces to a simple delta function when $\xi=0$. To see \eqref{eq:test_dist_1_3D} also satisfies its differential equation, namely \eqref{eq:LKFT_k12} for $\mathcal{K}_j$ with $j=1$ and $\epsilon=1/2$, which is explicitly written as
	\begin{align}
	& \quad \dfrac{\partial}{\partial\xi}\int ds\dfrac{\mathcal{K}_1(s,s';\xi)}{p^2-s+i\epsilon}\nonumber\\
	& =\alpha\mu\int ds\left[\dfrac{\sqrt{s}}{(p^2-s)^2}-\dfrac{\sqrt{s}}{2p^2(p^2-s)}-\dfrac{1}{2(p^2)^{3/2}}\mathrm{arctanh}(\sqrt{p^2/s}) \right]\mathcal{K}_1(s,s';\xi),\label{eq:LKFT_fermion_3D_k1}
	\end{align}
	We start with the following helpful relations, 
	\begin{align}
	& \int_{s_{th}}^{+\infty}ds\dfrac{1}{(p^2-s)s^{3/2}}=\dfrac{2}{(p^2)^{3/2}}\left[\sqrt{\dfrac{p^2}{s_{th}}}-\mathrm{arctanh}\sqrt{\dfrac{p^2}{s_{th}}} \right],\\[2mm]
	& \int_{s_{th}}^{+\infty}ds\left[\dfrac{1}{\sqrt{s}}-\dfrac{1}{\sqrt{p^2}}\mathrm{arctanh}\sqrt{\dfrac{p^2}{s}} \right]\dfrac{1}{s^{3/2}} =\dfrac{1}{s_{th}}-\dfrac{2}{\sqrt{s_{th}p^2}}\mathrm{arctanh}\sqrt{\dfrac{p^2}{s_{th}}}\nonumber\\
	& \hspace{8cm}-\dfrac{1}{p^2}\ln\left(1-\dfrac{p^2}{s_{th}}\right),
	\end{align}
	where $s_{th}=\left(\sqrt{s'}+\alpha\mu\xi/2\right)^2$. Next, applying \eqref{eq:test_dist_1_3D} and writing $1+\dfrac{\alpha\mu\xi}{2\sqrt{s'}}=\dfrac{\sqrt{s_{th}}}{\sqrt{s'}}$,
	\begin{align}
	& \quad \int ds\left[\dfrac{\sqrt{s}}{(p^2-s)^2}-\dfrac{\sqrt{s}}{2p^2(p^2-s)}-\dfrac{1}{2(p^2)^{3/2}}\mathrm{arctanh}(\sqrt{p^2/s}) \right]\mathcal{K}_1(s,s';\xi)\nonumber\\
	& =\left(1+\dfrac{\alpha\mu\xi}{2\sqrt{s'}}\right)^{-1}\left[\dfrac{\sqrt{s_{th}}}{(p^2-s_{th})^2}-\dfrac{1}{2\sqrt{s_{th}}(p^2-s_{th})}+\dfrac{1}{2p^2}\left(\dfrac{1}{\sqrt{s_{th}}}-\dfrac{1}{\sqrt{p^2}}\mathrm{arctanh}\sqrt{\dfrac{p^2}{s_{th}}} \right) \right]\nonumber\\
	&\quad +\dfrac{\alpha\mu\xi}{4}\Bigg\{\dfrac{-1}{s_{th}(p^2-s_{th})}+\dfrac{1}{s_{th}p^2}+\dfrac{1}{p^4}\ln\left(1-\dfrac{p^2}{s_{th}}\right)-\dfrac{1}{2}\left[\dfrac{1}{s_{th}p^2}+\dfrac{1}{p^4}\ln\left(1-\dfrac{p^2}{s_{th}}\right) \right]\nonumber\\
	& \quad +\dfrac{1}{2p^2}\left[\dfrac{1}{s_{th}}-\dfrac{1}{p^2}\ln\left(1-\dfrac{p^2}{s_{th}} \right)-\dfrac{2}{\sqrt{s_{th}p^2}}\mathrm{arctanh}\sqrt{\dfrac{p^2}{s_{th}}} \right] \Bigg\}\nonumber\\
	& =\dfrac{\sqrt{s'}}{(p^2-s_{th})^2}-\dfrac{1}{2\sqrt{s_{th}}(p^2-s_{th})}+\dfrac{1}{2\sqrt{s_{th}}p^2}-\dfrac{1}{2(p^2)^{3/2}}\mathrm{arctanh}\sqrt{\dfrac{p^2}{s_{th}}}.\label{eq:rhs_LKFT_fermion_3D_k1}
	\end{align}
	Meanwhile, since $\dfrac{\partial}{\partial\xi}\sqrt{s_{th}}=\dfrac{\alpha\mu}{2}$,
	\begin{align}
	&\quad \dfrac{\partial}{\partial\xi}\int ds\dfrac{1}{p^2-s}\mathcal{K}_1(s,s';\xi)\nonumber\\
	& =\dfrac{\partial}{\partial\xi}\left[\dfrac{\sqrt{s'}}{\sqrt{s_{th}}(p^2-s_{th})}+\dfrac{\alpha\mu\xi}{2p^2}\left(\dfrac{1}{\sqrt{s_{th}}}-\dfrac{1}{p^2}\mathrm{arctanh}\sqrt{\dfrac{p^2}{s_{th}}} \right) \right]\nonumber\\
	& =-\dfrac{\alpha\mu\sqrt{s'}}{2s_{th}(p^2-s_{th})}+\dfrac{\sqrt{s'}}{\sqrt{s_{th}}}\dfrac{\alpha\mu\sqrt{s_{th}}}{(p^2-s_{th})^2}\nonumber\\
	& \quad +\dfrac{\alpha\mu}{2p^2}\left(\dfrac{1}{\sqrt{s_{th}}}-\dfrac{1}{\sqrt{p^2}}\mathrm{arctanh}\sqrt{\dfrac{p^2}{s_{th}}} \right)-\dfrac{\alpha\mu\xi}{2p^2}\left(\dfrac{\alpha\mu}{2s_{th}}-\dfrac{1}{\sqrt{p^2}}\dfrac{\sqrt{p^2}\dfrac{\alpha\mu}{2s_{th}}}{1-\dfrac{p^2}{s_{th}}} \right)\nonumber\\
	& =\alpha\mu\left[\dfrac{\sqrt{s'}}{(p^2-s_{th})^2}-\dfrac{1}{2\sqrt{s_{th}}(p^2-s_{th})}+\dfrac{1}{2p^2\sqrt{s_{th}}}-\dfrac{1}{2(p^2)^{3/2}}\mathrm{arctanh}\sqrt{\dfrac{p^2}{s_{th}}} \right].\label{eq:lhs_LKFT_fermion_3D_k1}
	\end{align}
	The combination of \eqref{eq:lhs_LKFT_fermion_3D_k1} with \eqref{eq:rhs_LKFT_fermion_3D_k1} explicitly shows that $\mathcal{K}_1(s,s';\xi)$ given by \eqref{eq:test_dist_1_3D} indeed satisfies \eqref{eq:LKFT_fermion_3D_k1}. 
	\subsection{The closure property\label{ss:k1_3D_closure}}
	While for the group $\mathbf{K}$ defined by \eqref{eq:test_dist_1_3D},
	\begin{align}
	& \quad \int ds'\mathcal{K}_1(s,s';\xi)\mathcal{K}_1(s',s'';\xi')\nonumber\\
	& =\int ds'\Bigg\{ \left(1+\dfrac{\alpha\mu\xi}{2\sqrt{s'}}\right)^{-1}\delta\left(s-\left(\sqrt{s'}+\dfrac{\alpha\mu\xi}{2}\right)^2 \right)\nonumber\\
	& {\hspace{5.3cm}}\times\left(1+\dfrac{\alpha\mu\xi'}{2\sqrt{s''}}\right)^{-1}\delta\left(s'-\left(\sqrt{s''}+\dfrac{\alpha\mu\xi'}{2} \right)^2 \right) \nonumber\\
	& \quad +\left(1+\dfrac{\alpha\mu\xi}{2\sqrt{s'}}\right)^{-1}\delta\left(s-\left(\sqrt{s'}+\dfrac{\alpha\mu\xi}{2}\right)^2 \right)\dfrac{\alpha\mu\xi'}{4(s')^{3/2}}\theta\left(s'-\left(\sqrt{s''}+\dfrac{\alpha\mu\xi'}{2} \right)^2 \right)\nonumber\\
	& \quad +\dfrac{\alpha\mu\xi}{4s^{3/2}}\theta\left(s-\left(\sqrt{s'}+\dfrac{\alpha\mu\xi}{2}\right)^2 \right)\left(1+\dfrac{\alpha\mu\xi'}{2\sqrt{s''}}\right)^{-1}\delta\left(s'-\left(\sqrt{s''}+\dfrac{\alpha\mu\xi'}{2} \right)^2 \right)\nonumber\\
	& \quad +\dfrac{\alpha\mu\xi}{4s^{3/2}}\theta\left(s-\left(\sqrt{s'}+\dfrac{\alpha\mu\xi}{2}\right)^2 \right)\dfrac{\alpha\mu\xi'}{4(s')^{3/2}}\theta\left(s'-\left(\sqrt{s''}+\dfrac{\alpha\mu\xi'}{2} \right)^2 \right) \Bigg\}.\label{eq:Group_k1_3D}
	\end{align}
	Integrals for the first and third terms on the right-hand side of \eqref{eq:Group_k1_3D} are obvious. While for the second term, since $\sqrt{s'}>0$ and $\sqrt{s}>\alpha\mu\xi/2$
	\begin{equation}
	\delta\left(s-\left(\sqrt{s'}+\dfrac{\alpha\mu\xi}{2}\right)^2 \right)=\left(1+\dfrac{\alpha\mu\xi}{2\sqrt{s'}}\right)^{-1}\delta\left(s'-\left(\sqrt{s}-\dfrac{\alpha\mu\xi}{2}\right)^2 \right),
	\end{equation}
	and the theta function is not zero only when $\sqrt{s'}\geq \sqrt{s''}+\alpha\mu\xi'/2$. Therefore 
	\begin{align}
	& \quad \int ds'\left(1+\dfrac{\alpha\mu\xi}{2\sqrt{s'}}\right)^{-1}\delta\left(s-\left(\sqrt{s'}+\dfrac{\alpha\mu\xi}{2}\right)^2 \right)\dfrac{\alpha\mu\xi'}{4(s')^{3/2}}\,\theta\left(s'-\left(\sqrt{s''}+\dfrac{\alpha\mu\xi'}{2} \right)^2 \right)\nonumber\\
	& =\dfrac{\alpha\mu\xi'}{4s^{3/2}\left(1-\dfrac{\alpha\mu\xi}{2\sqrt{s}} \right)}\,\theta\left(s-\left[\sqrt{s''}+\dfrac{\alpha\mu}{2}(\xi+\xi') \right]^2 \right).
	\end{align}
	For the fourth term, two theta functions overlap only if $s\geq [\sqrt{s''}+\alpha\mu(\xi+\xi')/2]^2$. Then
	\begin{align}
	& \quad \dfrac{\alpha\mu\xi}{4s^{3/2}}\,\theta\left(s-\left(\sqrt{s'}+\dfrac{\alpha\mu\xi}{2}\right)^2 \right)\dfrac{\alpha\mu\xi'}{4(s')^{3/2}}\,\theta\left(s'-\left(\sqrt{s''}+\dfrac{\alpha\mu\xi'}{2} \right)^2 \right)\nonumber\\
	& =\theta\left(s-\left[\sqrt{s''}+\dfrac{\alpha\mu}{2}(\xi+\xi') \right]^2 \right)\int_{(\sqrt{s''}+\alpha\mu\xi'/2)^2}^{(\sqrt{s}-\alpha\mu\xi/2)^2}ds'\dfrac{(\alpha\mu)^2\xi\xi'}{16(ss')^{3/2}}\nonumber\\
	& =-\dfrac{\xi\xi'(\alpha\mu)^2}{8s^{3/2}}\left[\left(\sqrt{s}-\dfrac{\alpha\mu\xi}{2}\right)^{-1} -\left(\sqrt{s''}+\dfrac{\alpha\mu\xi'}{2}\right)^{-1}\right].
	\end{align}
	Therefore in the end,
	\begin{align}
	& \quad \int ds'\mathcal{K}_1(s,s';\xi)\mathcal{K}_1(s',s'';\xi')\nonumber\\
	& =\left[1+\dfrac{\alpha\mu}{2\sqrt{s''}}\left(\xi+\xi'\right) \right]^{-1}\delta\left(s-\left[\sqrt{s''}+\dfrac{\alpha\mu}{2}(\xi+\xi') \right]^2 \right)\nonumber\\
	& \quad +\theta\left(s-\left[\sqrt{s''}+\dfrac{\alpha\mu}{2}(\xi+\xi') \right]^2 \right) \Bigg\{\dfrac{\alpha\mu\xi'}{4s^{3/2}}\left(1-\dfrac{\alpha\mu\xi}{2\sqrt{s}} \right)^{-1}+\dfrac{\alpha\mu\xi}{4s^{3/2}}\left(1+\dfrac{\alpha\mu\xi'}{2\sqrt{s''}} \right)^{-1}\nonumber\\
	& \quad{\hspace{5cm}}  -\dfrac{\xi\xi'(\alpha\mu)^2}{8s^{3/2}}\left[\left(\sqrt{s}-\dfrac{\alpha\mu\xi}{2}\right)^{-1} -\left(\sqrt{s''}+\dfrac{\alpha\mu\xi'}{2}\right)^{-1}\right] \Bigg\}\nonumber\\
	& =\left[1+\dfrac{\alpha\mu}{2\sqrt{s''}}\left(\xi+\xi'\right) \right]^{-1}\delta\left(s-\left[\sqrt{s''}+\dfrac{\alpha\mu}{2}(\xi+\xi') \right]^2 \right)\nonumber\\
	& \quad{\hspace{5cm}}  +\dfrac{\alpha\mu(\xi+\xi')}{4s^{3/2}}\,\theta\left(s-\left[\sqrt{s''}+\dfrac{\alpha\mu}{2}(\xi+\xi') \right]^2 \right)\nonumber\\
	& =\mathcal{K}_1(s,s'';\xi+\xi').
	\end{align}
	So $\mathbf{K}$ defined by $\mathcal{K}_1(s,s';\xi)$ given by \eqref{eq:test_dist_1_3D} satisfies the closure property of a group.
\section*{References}
\bibliography{LKFT_fermion_epsilon_bib}
\end{document}